\documentclass[3p]{elsarticle}
\usepackage{lineno,hyperref}
\usepackage{amsmath}
\usepackage{color}
\usepackage{mathtools}
\usepackage[normalem]{ulem}
\usepackage{rotating}
\usepackage{makecell}
\usepackage{threeparttable, tablefootnote}

\journal{arXiv}
\newcommand{\Beta}{\mathrm{B}}

\begin{document}

\begin{frontmatter}

\title{Generalized Beta Prime Distribution: Stochastic Model of Economic Exchange and Properties of Inequality Indices}

\author[mymainaddress]{M. Dashti Moghaddam}
\author[mysecondaryaddress]{Jeffrey Mills\fnref{hisfootnote}}
\author[mymainaddress]{R. A. Serota\fnref{myfootnote}}
\fntext[myfootnote]{serota@ucmail.uc.edu}
\fntext[hisfootnote]{millsjf@ucmail.uc.eduu}

\address[mymainaddress]{Department of Physics, University of Cincinnati, Cincinnati, Ohio 45221-0011}
\address[mysecondaryaddress]{Department of Economics, University of Cincinnati, Cincinnati, Ohio 45221-0371}

\begin{abstract}
We argue that a stochastic model of economic exchange, whose steady-state distribution is a Generalized Beta Prime (also known as GB2), and some unique properties of the latter, are the reason for GB2's success in describing wealth/income distributions. We use housing sale prices as a proxy to wealth/income distribution to numerically illustrate this point. We also explore parametric limits of the distribution to do so analytically. We discuss parametric properties of the inequality indices -- Gini, Hoover, Theil T and Theil L -- vis-a-vis those of GB2 and introduce a new inequality index, which serves a similar purpose. We argue that Hoover and Theil L are more appropriate measures for distributions with power-law dependencies, especially fat tails, such as GB2.
\end{abstract}


\end{frontmatter}

\section{Introduction\label{Introduction}}

Voluminous literature exists \cite{chotikapanich2008modelling} on application of Generalized Beta Prime (also known as GB2) to wealth/income distributions. Analytical results are known (see Chapters 3 and 8 in \cite{mcdonald2008modelling}) for Theil and Gini measures of inequality. Numerical and analytical results using GB2 are routinely used in various economic analyses (for a recent example see \cite{chen2016influences, chotikapanich2018using}). Yet, there seems to be little discussion as to why GB2 is so uniquely suited for modeling wealth/income distributions. Additionally, there seems to be a glaring lack of maximum likelihood estimation (MLE) and Bayesian fitting, as well as of use of more accurate statistical measures of fits, such as Kolmogorov-Smirnov statistics. 

Consequently, the main motivation of this work is to connect GB2 with a plausible stochastic model \cite{hertzler2003classical} of economic exchange and to better understand parametric dependences and symmetries of the distribution and quantities derived thereof, such as Gini, Hoover (also known as Pietra or Schultz) and Theil indices. We also want to conduct a systematic statistical fitting and quantify goodness of fits. Finally, we propose an alternative measure of inequality, which -- similarly to Hoover and Theil L -- is less prone to exaggerating high and low income/wealth when described by power-law distribution dependencies, such as GB2.

This paper is organized as follows. In Section \ref{Distributions} we summarize the properties of distributions used for fitting in Section \ref{Simulations}. In Section \ref{Properties} we discuss the analytic properties of GB2 and inequality indices. Initially we discuss its particular case of Beta Prime, for which closed-form results are simple and easy to understand. In Section \ref{Stochastic} we discuss the stochastic model of economic exchange, whose steady-state distribution is GB2.  In Section \ref{Simulations} we use housing sales data as a proxy to wealth/income distribution for fitting with distributions of Section \ref{Distributions}. In \ref{New} we introduce a new measure of inequality, DMMS. In \ref{Expanded} we expand the number of fitting functions. Finally, in \ref{Tailcuts}-\ref{Market} we compare market values distribution with that of sale prices and investigate the effect of the most expensive properties on tail parameters. 

\section{Summary of Distributions \label{Distributions}}

Tables \ref{Analyticforms1} - \ref{Analyticforms3} contain probability density functions (PDF) and cumulative density function (CDF) of the distributions used for fitting in this article and inequality indices derived for them.\footnote{BP is a particular case of GB2. Expanded tables, that contain IGa and Ga -- particular cases of GIGa and GGa -- are given in \ref{Expanded}.} Here $\Gamma (x)$ is the gamma function, $\Beta(x,y)$ is the beta function, $\psi(x)$ is the digamma function, $erf(x)$ is the error function and $\prescript{}{m}{F}{}_{n}$ is the hypergeometric function. The fat-tailed Generalized Inverse Gamma (GIGa) appear in network models of economy \cite{bouchaud2000wealth,ma2013distribution} and is related to Generalized Gamma (GGa) by changing the variable of the PDF to its inverse. These distributions and their particular cases also appear in the models of stochastic volatility and stock returns \cite{dragulescu2002probability, ma2014model}; for the application of BP in the the latter context see \cite{dashti2018combined}. GIGa and GGa can be also viewed as the limiting cases of GB2 for $p \rightarrow \infty$ and $q \rightarrow \infty$ respectively \cite{mcdonald1995generalization}. Lognormal (LN) distribution is also widely used in economics and finance \cite{limpert2001lognormal}. 

A succinct summary of inequality indices for many distributions can be found in \cite{mcdonald2008modelling}. We derived all inequality indices for GIGa, as well as Theil L and Hover for GB2, for this paper. However, we subsequently found that a very nice, closed-form formula for the Hoover index (Prieta there) had been previously obtained for an arbitrary distribution in terms of its CDF \cite{sarabia2014explicit} and that Theil L had also been previously derived  \cite{chotikapanich2018using}. We point out the difference structure of all the indices which reflect the symmetry associated with the transformation of variable of the distribution to its inverse, which converts low end of GB2 to high end and vice versa (see below) and GIGa to GGa and vice versa. 

Gini, Hoover and Theil indices are calculated using the following formulae, where $p(x)$ is a PDF:
\begin{equation}
G =\frac{1}{2 \mu} \int_{0}^{\infty} \int_{0}^{\infty} p(x) p(y) |x - y| \mathrm{d}x \mathrm{d}y
\label{Gini}
\end{equation}

\begin{equation}
H =\frac{1}{2 \mu}\int_{0}^{\infty} p(x) |x - \mu |  \mathrm{d}x 
\label{Hoover}
\end{equation}

\begin{equation}
T_{T} =\int_{0}^{\infty} p(x) \frac{x}{\mu} \ln{\frac{x}{\mu}}  \mathrm{d}x 
\label{TheilT}
\end{equation}

\begin{equation}
T_{L} =\int_{0}^{\infty} p(x) \ln{\frac{\mu}{x}}  \mathrm{d}x 
\label{TheilL}
\end{equation}
From (\ref{Gini}) - (\ref{TheilL}), it is clear that Gini and Theil T exaggerate power-law dependences of a distribution -- low-end for GGa, high-end fat tails for GIGa and both for GB2 (and BP) -- in that they underweigh low end and overweigh fat tails. Consequently, we believe that Hoover and Theil L are more appropriate measures for distributions with power laws, especially with fat tails. In \ref{New} we introduce yet another measure of inequality, DMMS, which tries to address the same problem. As will be seen below, Theil L is smaller than Theil T both for actual data and for fat-tailed distribution fits, while Hoover and MDDS are smaller than Gini, with MDDS being smaller than Hoover for all fitted distributions.

\begin{sidewaystable}
\centering
\caption{Analytic Forms of Distributions}
\label{Analyticforms1}
\fontsize{7.5}{8}
\begin{tabular}{|c|c|c|c|} 
\hline
            type &       PDF &        CDF &          Mean ($\mu$)  \\
\hline
GB2 & $\frac{\alpha(1+(\frac{x}{\beta})^{\alpha})^{-p-q}(\frac{x}{\beta})^{-1+p\alpha}}{\beta \space \Beta(p,q)}$ &$I(\frac{x^\alpha}{x^\alpha+\beta^\alpha},p,q)$ &  $\frac{\beta \Beta(p + \frac{1}{\alpha},q - \frac{1}{\alpha})}{\Beta(p,q)}$\\
\hline
BP & $ \frac{(1+\frac{x}{\beta})^{-p-q}(\frac{x}{\beta})^{-1+p}}{\beta \space B(p,q)}$ &  $I(\frac{x}{x+\beta},p,q)$&$\frac{\beta p}{q -1}$ \\
\hline
GIGa & $\frac{\gamma e^{-(\frac{\beta}{x})^\gamma}(\frac{\beta}{x})^{1+\alpha\gamma}}{\beta \Gamma(\alpha)}$& $Q(\alpha,(\frac{\beta}{x})^\gamma) $ &$\frac{\beta \Gamma(\alpha - \frac{1}{\gamma})}{\Gamma(\alpha)}$  \\
\hline
LN & $\frac{1}{x \sigma \sqrt{2 \pi}} e^{-(\frac{\ln x - \mu}{2 \sigma ^ 2})}$& $ \frac{1}{2} + \frac{1}{2} erf(\frac{\ln x - \mu}{\sqrt{2} \sigma})$& $e^{\mu+\frac{\sigma^2}{2}}$\\
\hline
GGa & $\frac{\gamma e^{-(\frac{x}{\beta})^{\gamma}}(\frac{x}{\beta})^{-1+\alpha \gamma}}{ \beta \Gamma(\alpha)}$ & $1 - Q(\alpha,(\frac{x}{\beta})^\gamma)$ & $\frac{\beta \Gamma(\alpha + \frac{1}{\gamma})}{\Gamma(\alpha)}$  \\
\hline
\hline
\end{tabular}

    \vspace{2\baselineskip}
    \centering
\caption{Gini and Hoover Indices}
\label{Analyticforms2}
\fontsize{7.5}{8}
\begin{tabular}{|c|c|c|} 
\hline
            type &       Gini &        Hoover  \\
\hline
GB2 &  \makecell{$\frac{\Beta(2q-\frac{1}{\alpha},2p+\frac{1}{\alpha})}{\Beta(p,q) \Beta(p+\frac{1}{\alpha},q-\frac{1}{\alpha})} (\frac{1}{p} \prescript{}{3}{F}{}_{2}(1,p+q,2p+\frac{1}{\alpha};p+1,2(p+q);1)-$  \\$\frac{1}{p+\frac{1}{\alpha}}\prescript{}{3}{F}{}_{2}(1,p+q,2p+\frac{1}{\alpha};p+1+\frac{1}{\alpha},2(p+q);1))$} & $I(\frac{(\frac{\mu}{\beta})^{\alpha}}{1+(\frac{\mu}{\beta})^{\alpha}},p,q)- I(\frac{(\frac{\mu}{\beta})^{\alpha}}{1+(\frac{\mu}{\beta})^{\alpha}},p+\frac{1}{\alpha},q-\frac{1}{\alpha})$ \\
\hline
BP &  \makecell{$\frac{\Beta(2p+1,2q-1)}{\Beta(p,q) \Beta(p+1,q-1)} (\frac{1}{p} \prescript{}{3}{F}{}_{2}(1,p+q,2p+1;p+1,2(p+q);1)-$  \\$\frac{1}{p+1}\prescript{}{3}{F}{}_{2}(1,p+q,2p+1;p+2,2(p+q);1))$\\$=\frac{\Beta(2p + 1,2q-1)}{\Beta(p,q) \Beta(p+1,q-1)}\frac{2p+2q-1}{p(q-1)}$} &$I(\frac{\frac{\mu}{\beta}}{1+\frac{\mu}{\beta}},p,q) - I(\frac{\frac{\mu}{\beta}}{1+\frac{\mu}{\beta}},p+1,q-1) = \frac{p^{-1 + p} (-1 + q)^{-1 + q} (-1 + p + q)^{1 - p - q}}{\Beta(p,q)}$ \\
\hline
GIGa &  \makecell{$\frac{1}{\Beta(\alpha,\alpha-\frac{1}{\gamma})}(\frac{1}{\alpha-\frac{1}{\gamma}} \prescript{}{2}{F}{}_{1}(\alpha-\frac{1}{\gamma},2 \alpha-\frac{1}{\gamma};\alpha-\frac{1}{\gamma}+1;-1) - $\\$(\frac{1}{\alpha}) \prescript{}{2}{F}{}_{1}(\alpha,2 \alpha-\frac{1}{\gamma};\alpha + 1;-1))$} &  $Q(\alpha,(\frac{\beta}{\mu})^\gamma) - Q(\alpha- \frac{1}{\gamma},(\frac{\beta}{\mu})^\gamma)$\\
\hline
LN & $erf (\frac{\sigma}{2})$  & $  erf (\frac{\sigma}{ 2 \sqrt{2}}) $\\
\hline
GGa & \makecell{$\frac{1}{2^{2 \alpha + \frac{1}{\gamma}} \Beta(\alpha,\alpha+\frac{1}{\gamma})}(\frac{1}{\alpha} \prescript{}{2}{F}{}_{1}(1,2 \alpha+\frac{1}{\gamma};\alpha+1;\frac{1}{2}) - $\\$(\frac{1}{\alpha+\frac{1}{\gamma}}) \prescript{}{2}{F}{}_{1}(1,2 \alpha+\frac{1}{\gamma};\alpha + \frac{1}{\gamma};\frac{1}{2}))
$ } & $ Q(\alpha + \frac{1}{\gamma},(\frac{\mu}{\beta})^\gamma)- Q(\alpha,(\frac{\mu}{\beta})^\gamma)$\\
\hline
\hline
\end{tabular}

    \vspace{2\baselineskip}
    \centering
\caption{Theil T and Theil L Indices}
\label{Analyticforms3}
\fontsize{7.5}{8}
\begin{tabular}{|c|c|c|} 
\hline
            type &  Theil T &Theil L \\
\hline
GB2 &  $\frac{1}{\alpha}(\psi(p+\frac{1}{\alpha})-\psi(q-\frac{1}{\alpha}))+\ln \frac{B(p,q)}{B(p+\frac{1}{\alpha},q-\frac{1}{\alpha})}$ & \makecell{$\frac{1}{\alpha}(\psi(q)-\psi(p))-$\\$\ln \frac{B(p,q)}{B(p+\frac{1}{\alpha},q-\frac{1}{\alpha})}$} \\
\hline
BP &    $(\psi(p + 1)-\psi(q - 1)) + \ln \frac{q-1}{p}$   &  $(\psi(q)-\psi(p)) - \ln \frac{q-1}{p}$ \\
\hline
GIGa &    $-\frac{1}{\gamma}\psi(\alpha - \frac{1}{\gamma}) + \ln \frac{\Gamma(\alpha)}{\Gamma(\alpha-\frac{1}{\gamma})}$       &     $\frac{1}{\gamma}\psi(\alpha) - \ln \frac{\Gamma(\alpha)}{\Gamma(\alpha-\frac{1}{\gamma})}$\\
\hline
LN &  $\frac{\sigma^2}{2}$  &  $\frac{\sigma^2}{2}$  \\
\hline
GGa &  $\frac{1}{\gamma}\psi(\alpha + \frac{1}{\gamma}) + \ln \frac{\Gamma(\alpha)}{\Gamma(\alpha+\frac{1}{\gamma})}$  &  $-\frac{1}{\gamma}\psi(\alpha) + \ln\frac{\Gamma(\alpha+\frac{1}{\gamma})}{\Gamma(\alpha)}$\\
\hline
\hline
\end{tabular}
\begin{tablenotes}
  \item[*] $B(p,q) = \frac{\Gamma(p)\Gamma(q)}{\Gamma(p+q)}$: beta function; $\Gamma(\alpha)$: gamma function.
  \item[a] $\psi(x) = \frac{\mathrm{d}\ln\Gamma(x)}{\mathrm{d}x} = \frac{\Gamma'(x)}{\Gamma(x)}$: digamma function.
  \item[d] $Q(\alpha,x)= \frac{\Gamma(\alpha,x)}{\Gamma(\alpha)}$: regularized gamma function; $\Gamma(\alpha,x)$: incomplete gamma function
   \item[e] $I(x,p,q)=\frac{\Beta(x,p,q)}{\Beta(p,q)}$: regularized beta function; $\Beta(x,p,q)$: incomplete beta function
  \end{tablenotes}
   \end{sidewaystable}

\section{Analytic Properties of Generalized Beta and Inequality Indices \label{Properties}}

\subsection {Beta Prime\label{Beta}}
We begin our analysis of GB2
\begin{equation}
GB2(x; p,q,\beta,\alpha)=\frac{\alpha (1+({\frac{x}{\beta}})^{\alpha})^{-p-q}(\frac{x}{\beta})^{-1+p\alpha}}{\beta \Beta(p,q)}
\label{GB2}
\end{equation}
with the simpler case of BP, $\alpha=1$ in (\ref{GB2}).
\begin{equation}
BP(x; p,q,\beta)=\frac{(1+({\frac{x}{\beta}}))^{-p-q}(\frac{x}{\beta})^{-1+p}}{\beta \Beta(p,q)}
\label{BP}
\end{equation}
with the simpler case of BP, $\alpha=1$ in (\ref{GB2}). The limiting behaviors of BP at small and large arguments are 
\begin{equation}
BP(x; p,q,\beta) \propto (\frac{x}{\beta})^{-q-1}, \hspace{.25cm} x \gg \beta
\label{BPlargex}
\end{equation}
and
\begin{equation}
BP(x; p,q,\beta) \propto (\frac{x}{\beta})^{p-1}, \hspace{.25cm} x \ll \beta
\label{BPsmallx}
\end{equation}
Here we assume that $p>1$, that is that BP has a maximum (a bell shape), and $q>2$, that is that the variance exists (we will discuss the example to the contrary for market values in the Appendix). The limiting behaviors (\ref{BPlargex}) and (\ref{BPsmallx}) underscore the flexibility of BP: large-$p$ behavior mimics exponential decay of IGa (see \ref{Expanded}) at small values and corresponds to suppression of low-end values, while large-$q$ behavior mimics exponential decay of Ga and corresponds to suppression of high-end values (see also \cite{mcdonald1995generalization}).

An important property of BP is that under the change of variable to its inverse, $x \to x^{-1}$, which converts low end to high end and vice versa, it transforms as \footnote{Notice that under such transformation $LN \to LN$, $Ga \to IGa$ and $IGa \to Ga$}
\begin{equation}
BP(x; p,q,\beta) \to BP(x^{-1}; q,p,\beta^{-1})
\label{BPtransform}
\end{equation}
From definitions of Gini and Theil it can be then derived -- and immediately verified from the closed form answers in Tables \ref{Analyticforms2} - \ref{Analyticforms3} -- that the transformation property (\ref{BPtransform}) leads to $p+1 \leftrightarrow q$ transformation property for the inequality indices:
\begin{equation}
G^{BP} (p,q) = G^{BP} (q-1,p+1); H^{BP} (p,q) = H^{BP} (q-1,p+1)
\label{GiniBPtransform}
\end{equation}
and
\begin{equation}
\begin{multlined}
T^{BP}_{T} (p,q) = T^{BP}_{L} (q-1,p+1) \\*
T^{BP}_{L} (p,q) = T^{BP}_{T} (q-1,p+1)
\end{multlined}
\label{TheilPBtransform}
\end{equation}
Notice that $p+1 \leftrightarrow q$ implies that $p>1$ and $q>2$ transform properly as well, which supports these conditions for BP. It is said that Theil L is more sensitive to inequality at low end -- due to the $ln(x)$, for $x$ in units of mean, which diverges as $x \to 0$ -- while Theil T is more sensitive to inequality at high end -- due to the $xln(x)$, which diverges as $x \to \infty$. We believe that (\ref{TheilPBtransform}) is a more precise formulation of their roles.

\begin{figure}[!htbp]
\centering
\begin{tabular}{cc}
\includegraphics[width = 0.49 \textwidth]{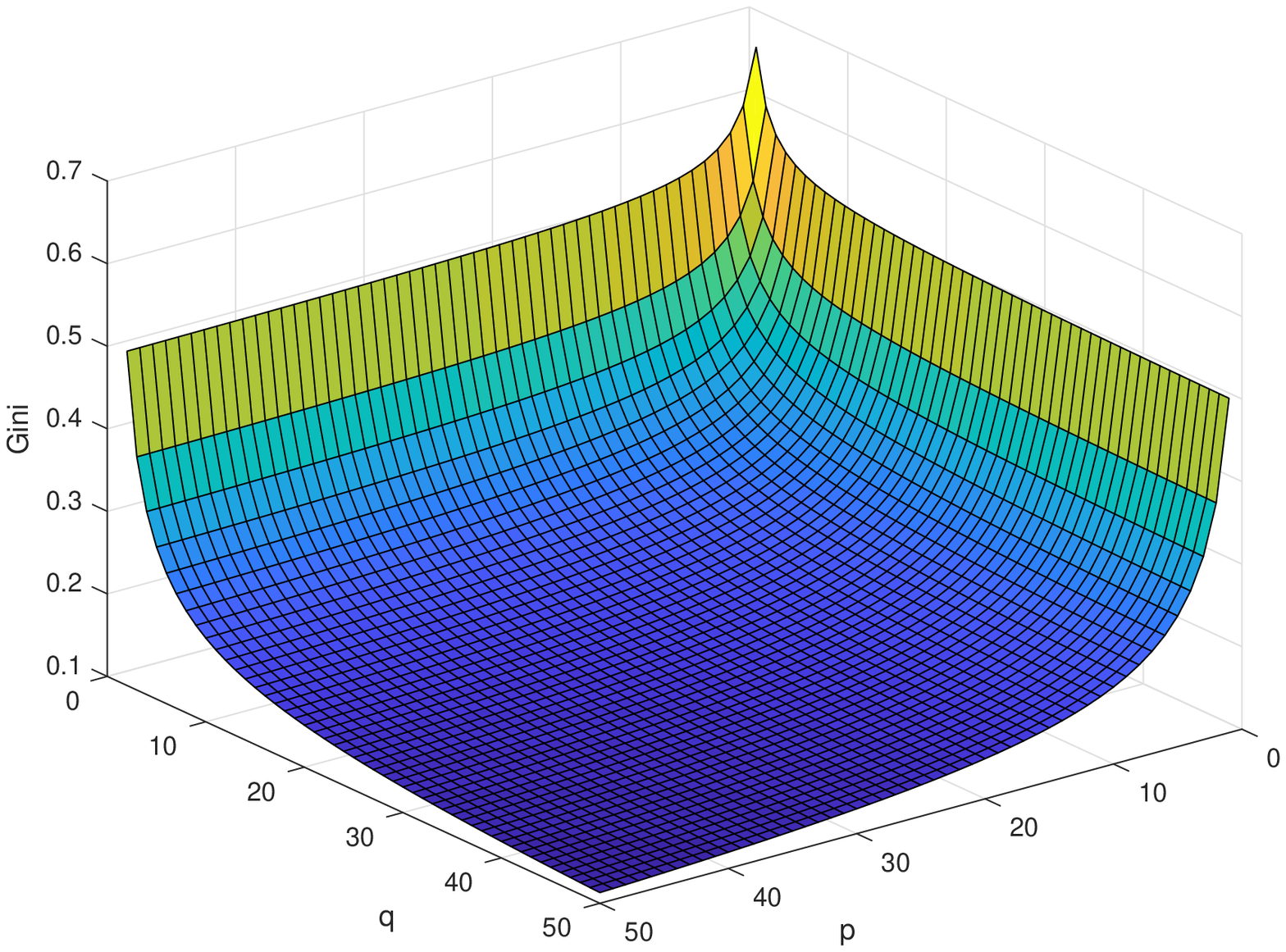}
\includegraphics[width = 0.49 \textwidth]{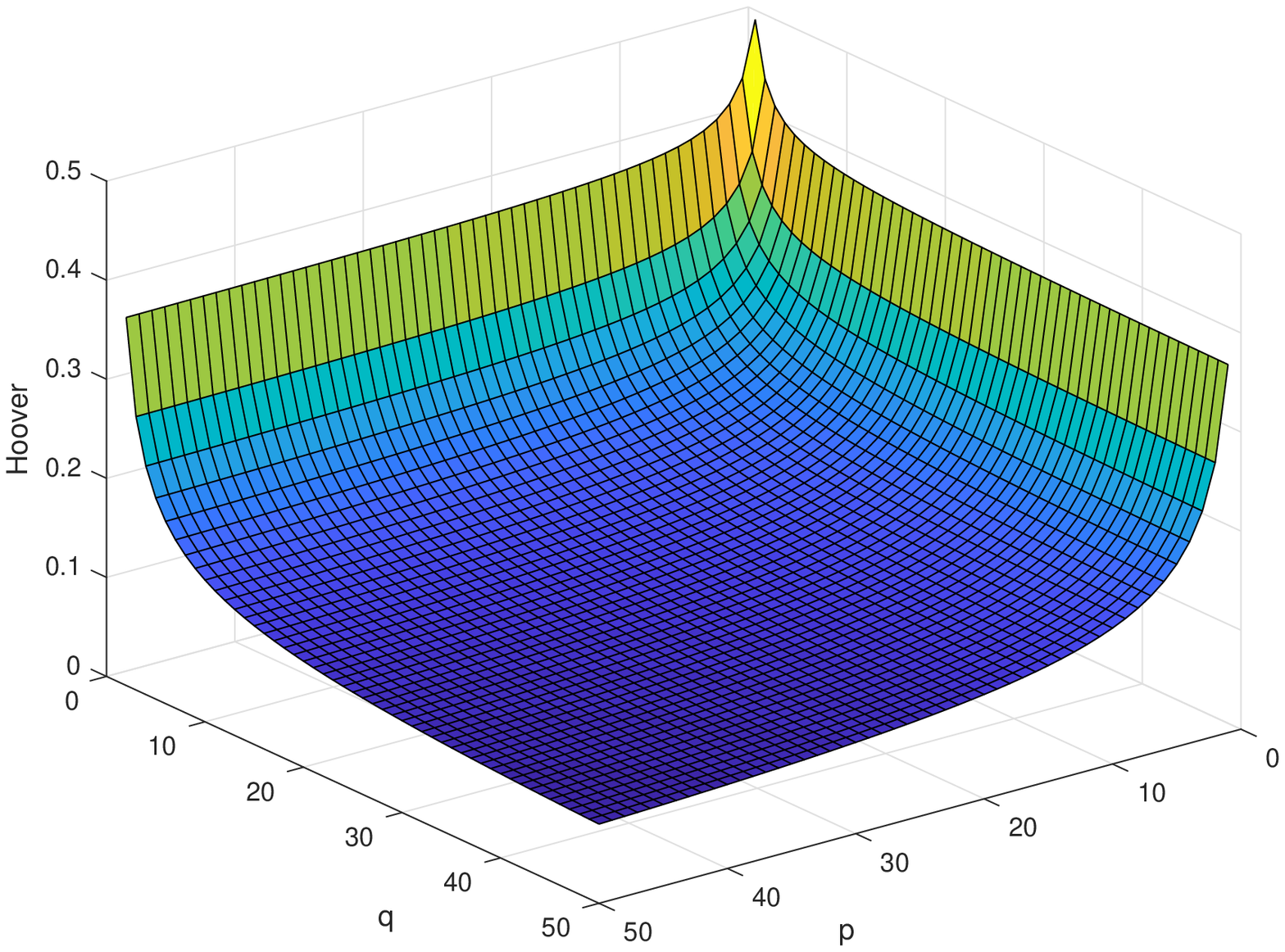}
\end{tabular}
\caption{Beta Prime Gini (left) and Hoover (right) as a function of $p$ and $q$, (\ref{GiniBP}.)}
\label{Gini3D}
\end{figure}

\begin{figure}[!htbp]
\centering
\begin{tabular}{cc}
\includegraphics[width = 0.49 \textwidth]{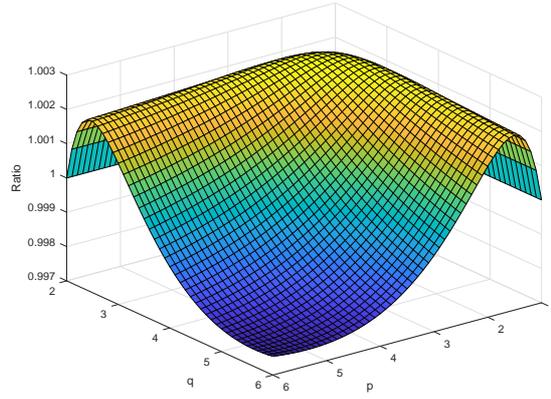}
\end{tabular}
\caption{Ratio of approximate expression for Gini, (\ref{GiniBPapprox}), to exact Gini, (\ref{GiniBP}): $\frac {\frac{pq+6p+7q-6}{8(pq+q-1)}}{\frac{2 \Beta(2p,2q-1)}{p \Beta^2(p,q)}}$.}
\label{ratio}
\end{figure}

We now turn to some limiting behaviors of Gini and Theil. As per Table \ref{Analyticforms3},
\begin{equation}
G^{BP} = \frac{2 \Beta(2p,2q-1)}{p \Beta^2(p,q)} ; H^{BP} = \frac{p^{-1 + p} (-1 + q)^{-1 + q} (-1 + p + q)^{1 - p - q}}{\Beta(p,q)}
\label{GiniBP}
\end{equation}
(Notice that this expression for $G^{BP}$ \cite{mcdonald2008modelling} is equivalent to the one in Table \ref{Analyticforms2}; the latter is written in a way to underscore $p+1 \leftrightarrow q$ symmetry.) Given the imposed constraints $p>1$ and $q>2$, the maximum value of Gini in BP is
\begin{equation}
G^{BP}_{max}=G^{BP} (1,2)=\frac{2}{3} ; H^{BP}_{max}=H^{BP} (1,2)= \frac{1}{2}
\label{GiniBPmax}
\end{equation}
Also notably
\begin{equation}
\begin{multlined}
G^{BP} (p,q \gg 1) =\frac{\Gamma(p+\frac{1}{2})}{\sqrt{\pi}\Gamma(p+1)} \\*
G^{BP} (p \gg 1,q) =\frac{\Gamma(q-\frac{1}{2})}{\sqrt{\pi}\Gamma(q)}
\end{multlined}
\begin{multlined}
; H^{BP} (p,q \gg 1) =  \frac{e^{-p} p^{p-1}}{\Gamma (p)}\\*
; H^{BP} (p \gg 1,q) = \frac{e^{1-q} (q-1)^{q-1}}{\Gamma (q)}
\end{multlined}
\label{GiniPBlimits}
\end{equation}
and, in particular,
\begin{equation}
\begin{multlined}
G^{BP} (1,q \gg 1) = \frac{1}{2} + \frac{1}{4q}\\*
G^{BP} (p \gg 1,2) =\frac{1}{2} + \frac{1}{4p}
\end{multlined}
\begin{multlined}
; H^{BP} (1,q \gg 1) =\frac{1}{e} + \frac{1}{2 e q}\\*
; H^{BP} (p \gg 1,2) =\frac{1}{e} + \frac{1}{2 e p}
\end{multlined}
\label{GiniPBlimits12}
\end{equation}
When both parameters become large, Gini tends to zero as 
\begin{equation}
G^{BP} (p \gg 1, q \gg 1) \approx \frac{1}{\sqrt{2\pi}}(\frac{1}{\sqrt{p}}+\frac{1}{\sqrt{q}}) ; H^{BP} (p \gg 1, q \gg 1) \approx \frac{1}{\sqrt{2\pi}}(\frac{1}{\sqrt{p}}+\frac{1}{\sqrt{q}})
\label{GiniPBlimitsboth}
\end{equation}
The dependence of BP Gini on $p$ and $q$ is summarized in Fig. \ref{Gini3D}. The slight asymmetry in dependences underscores the aforementioned $p+1 \leftrightarrow q$ symmetry and disappears for large $p$ and $q$, as per (\ref{GiniPBlimits}) -- (\ref{GiniPBlimitsboth}). Finally, while (\ref{GiniBP}) is simple enough and can be easily tabulated, a much simpler expression
\begin{equation}
G^{BP} (p,q) \approx \frac{pq+6p+7q-6}{8(pq+q-1)}
\label{GiniBPapprox}
\end{equation}
is within a fraction of a percentage point of exact value for relatively small $p$ and $q$ that are usually of interest, such as in Section \ref{Simulations}; (\ref{GiniBPapprox}) also respects $p+1 \leftrightarrow q$ symmetry. The ratio of expression in (\ref{GiniBPapprox}) to exact BP Gini in (\ref{GiniBP}) is shown in Fig. \ref{ratio}.

\begin{figure}[!htbp]
\centering
\begin{tabular}{cc}
\includegraphics[width = 0.49 \textwidth]{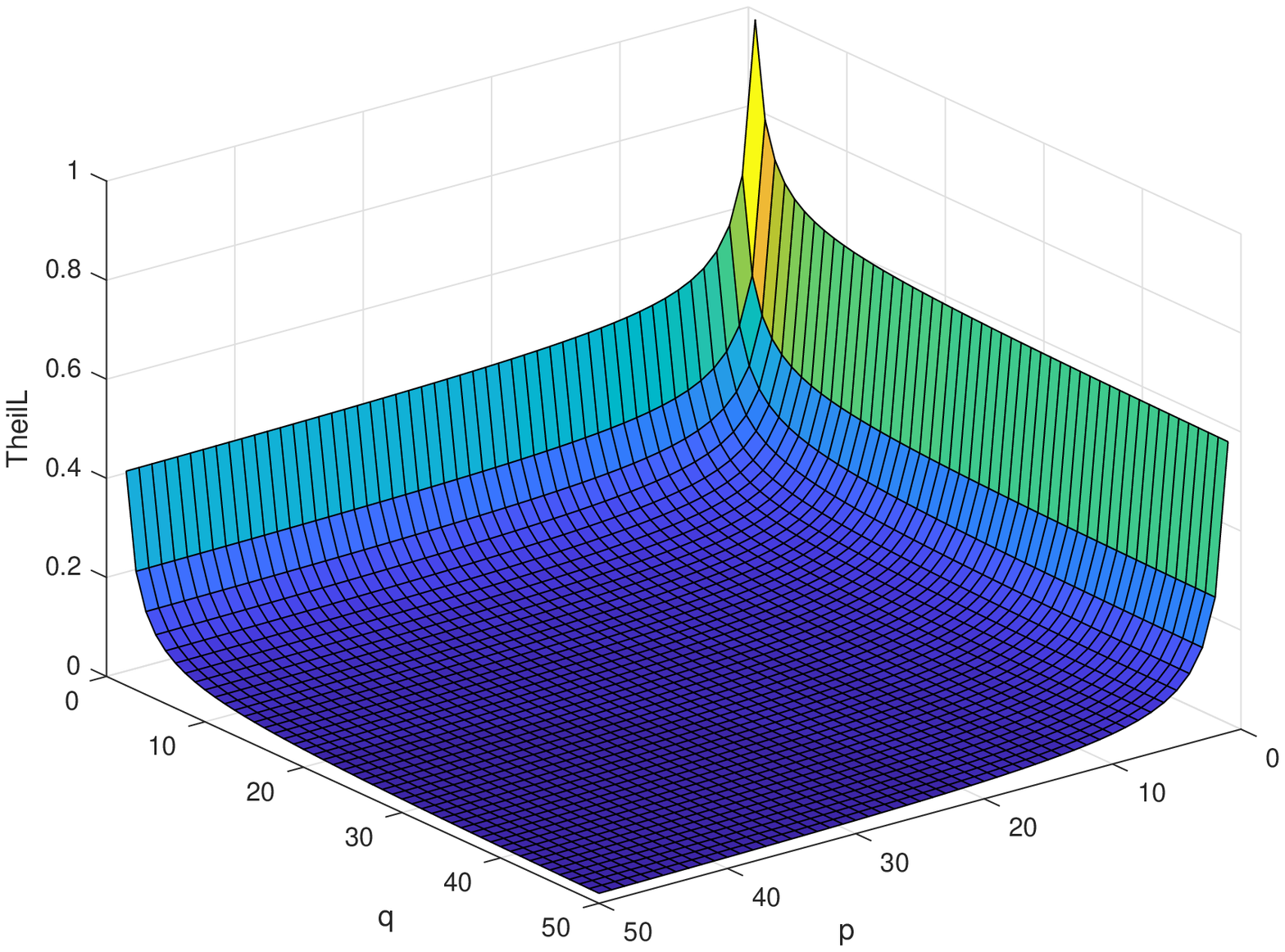}
\includegraphics[width = 0.49 \textwidth]{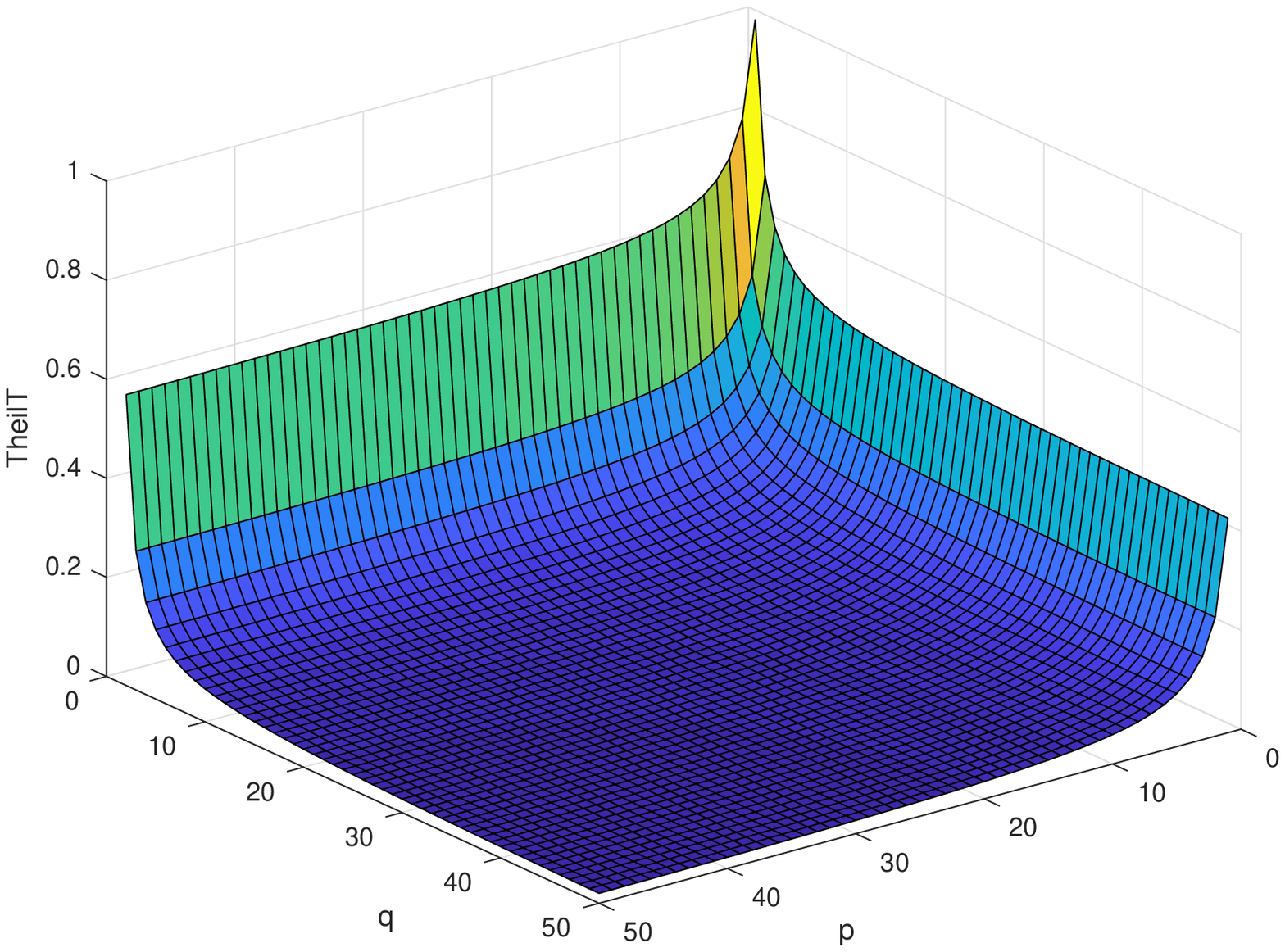}
\end{tabular}
\caption{Theil T (left) and Theil L (right), (\ref{TheilBP}).}
\label{Theils3D}
\end{figure}

\begin{figure}[!htbp]
\centering
\begin{tabular}{cc}
\includegraphics[width = 0.49 \textwidth]{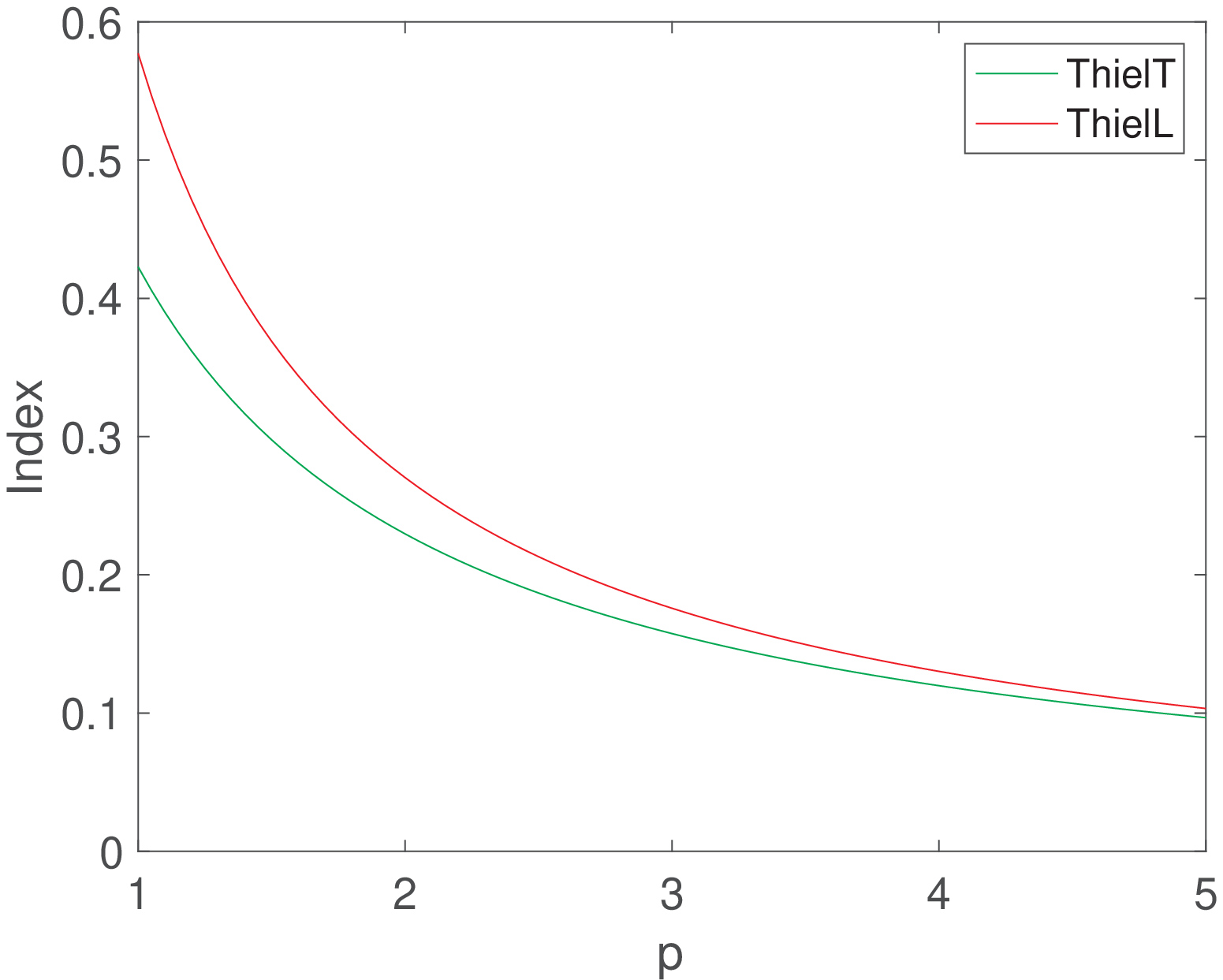}
\includegraphics[width = 0.49 \textwidth]{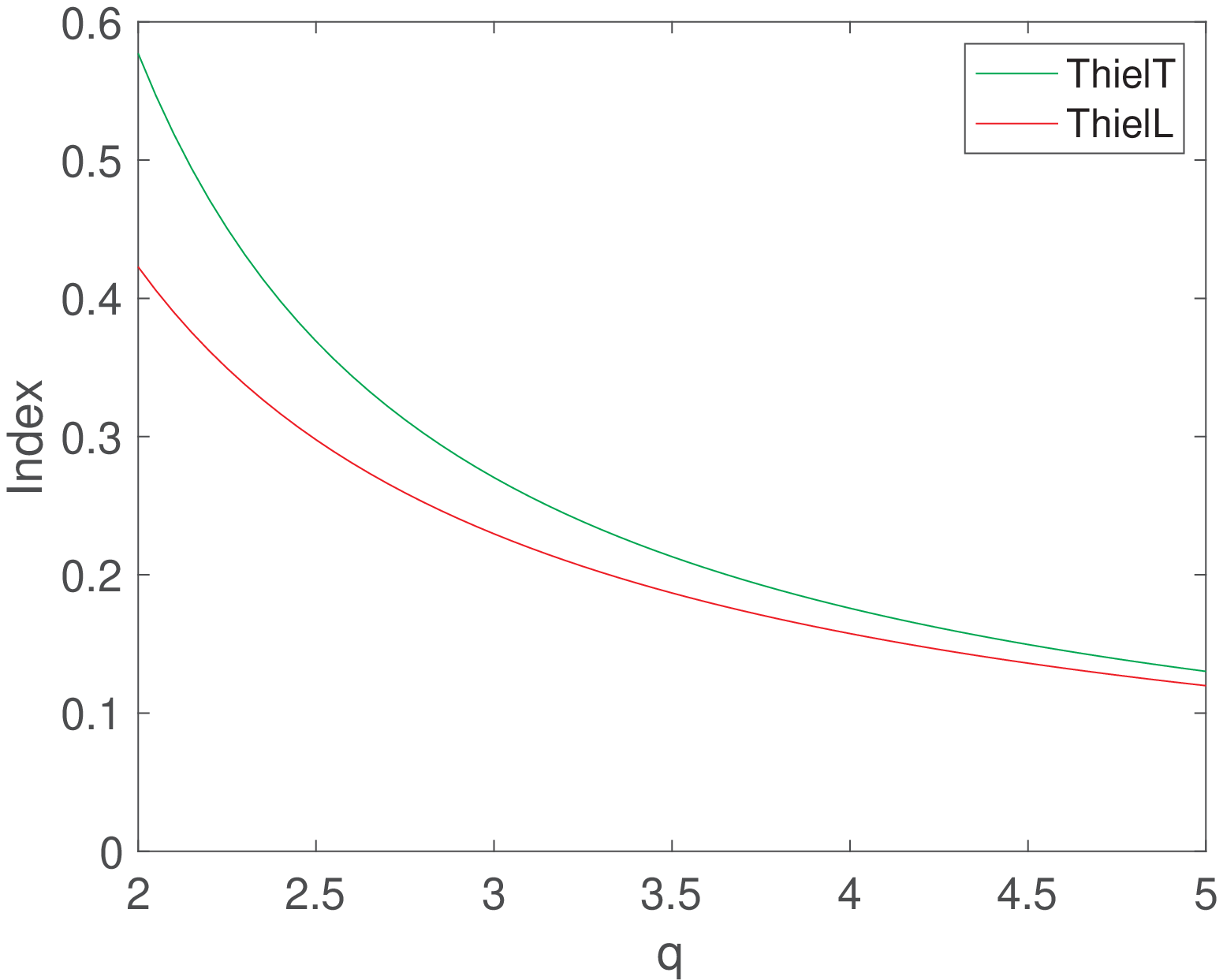}
\end{tabular}
\caption{Per (\ref{TheilBP}). Left: $p$-dependence of Theil T (top line, red), $\psi(p+1)-\ln p$, and Theil L (bottom line, green),  $-\psi(p)+\ln p$. Right: $q$-dependence of Theil T (bottom line, red), $- \psi(q-1) + \ln (q-1)$, and Theil L (top line, green),  $ \psi(q) - \ln (q-1)$. }
\label{Theil_qp}
\end{figure}

We now turn to Theil. As per Table \ref{Analyticforms3}, 
\begin{equation}
\begin{multlined}
T^{BP}_{T} (p,q)=\psi(p + 1)-\psi(q - 1) + ln(\frac{q-1}{p}) = \psi(p+1)-\ln p - \psi(q-1) + \ln (q-1)\\*
T^{BP}_{L} (p,q)=\psi(q)-\psi(p)) - ln(\frac{q-1}{p}) = -\psi(p)+\ln p + \psi(q) - \ln (q-1))
\end{multlined}
\label{TheilBP}
\end{equation}
Both Theil T and Theil L are shown in Fig. \ref{Theils3D}. The mirror reflection between the two are as per (\ref{TheilPBtransform}) and the slight asymmetry between $p$ and $q$ is per $p+1 \leftrightarrow q$, as was the case for Gini. Additionally,
\begin{equation}
T^{BP}_{T,max} = T^{BP}_{L,max} = T^{BP}_{T,L} (1,2) =1
\label{TheilBPmax}
\end{equation}
and for large $p$ and $q$ we just note that
\begin{equation}
\begin{multlined}
T^{BP}_T (1,q \gg 1) =1-\gamma+\frac{1}{2q} \\*
T^{BP}_T (p \gg 1,2) =\gamma+\frac{1}{2p} \\*
T^{BP}_T (p \gg 1,q \gg 1) =\frac{1}{2q}+\frac{1}{2p} 
\end{multlined}
\label{TheilPBlimits}
\end{equation}
where $\gamma = - \psi(1) \approx 0.577$ is Euler's gamma. For $p,q \gg 1$, the results for $T^{BP}_L (p,q)$ are obtained via $p \leftrightarrow q$.
Notice that (\ref{TheilBP}) indicates that Theil's dependence on $p$ and $q$ decouples, as shown in Fig. \ref{Theil_qp}.

\subsection {Generalized Beta Prime\label{Generalized}}
We now turn to the properties of GB2 distribution, whose PDF is given by (\ref{GB2}). Its limiting behaviors are given by 
\begin{equation}
GB2(x; p,q,\alpha,\beta) \propto (\frac{x}{\beta})^{-\alpha q -1}, \hspace{.25cm} x \gg \beta
\label{GB2largex}
\end{equation}
and
\begin{equation}
GB2(x; p,q,\alpha,\beta) \propto (\frac{x}{\beta})^{\alpha p -1}, \hspace{.25cm} x \ll \beta
\label{GB2smallx}
\end{equation}
Here we again assume that $\alpha p>1$, that is that BP has a maximum (a bell shape), and $\alpha q>2$, that is that the variance exists. Just as for BP, the important property of GB2 is that under the change of variable to its inverse, $x \to x^{-1}$, which converts low end to high end and vice versa, it transforms as
\begin{equation}
GB2 (x; p, q, \alpha, \beta) \to GB2(x^{-1}; q, p, \alpha, \beta^{-1})
\label{GB2transform}
\end{equation}
Sequentially, similarly to BP, we recognize the symmetry $p+\frac{1}{\alpha} \leftrightarrow q$, which leads to the following relationships for the Gini coefficient
\begin{equation}
G^{GB2} (p,q) = G^{GB2} (q-\frac{1}{\alpha},p+\frac{1}{\alpha}) ; H^{GB2} (p,q) = H^{GB2} (q-\frac{1}{\alpha},p+\frac{1}{\alpha})
\label{GiniGB2transform}
\end{equation}
and for Theil coefficients
\begin{equation}
\begin{multlined}
T^{GB2}_{T} (p,q) = T^{GB2}_{L} (q-\frac{1}{\alpha},p+\frac{1}{\alpha})= \frac{1}{\alpha}\psi(p+\frac{1}{\alpha})-\ln\frac{\Gamma(p+\frac{1}{\alpha}))}{\Gamma(p)}-\frac{1}{\alpha}\psi(q-\frac{1}{\alpha}) + \ln\frac{\Gamma(q))}{\Gamma(q-\frac{1}{\alpha})} \\*
T^{GB2}_{L} (p,q) = T^{GB2}_{T} (q-\frac{1}{\alpha},p+\frac{1}{\alpha})= -\frac{1}{\alpha}\psi(p)+\ln\frac{\Gamma(p+\frac{1}{\alpha}))}{\Gamma(p)}+\frac{1}{\alpha}\psi(q) - \ln\frac{\Gamma(q))}{\Gamma(q-\frac{1}{\alpha})}
\end{multlined}
\label{TheilGB2transform}
\end{equation}
where we again decoupled dependence on $p$ and $q$. While the $p+\frac{1}{\alpha} \leftrightarrow q$ symmetry for Hoover and between $T_T$ and $T_L$ is easily verified analytically from Table \ref{Analyticforms3} and (\ref{TheilGB2transform}), we were able to verify one for Gini only numerically.

\section{Stochastic Model of Economic Exchange \label{Stochastic}}

We now discuss the stochastic model of economic exchange that may be an underlying cause for the GB2 wealth/income distribution. \footnote{Recently, it was proposed that GB2 may also describe market volatility \cite{dashti2019implied, yan2019general}, which makes this model even more relevant.} As before, we begin with the BP discussion, which is very transparent an easy to understand.

\subsection{Beta Prime SDE}

The mean-reverting stochastic differential equation (SDE), whose steady-state distribution is given by BP can be written as \cite{hertzler2003classical, dashti2018combined} 
\begin{equation}
\mathrm{d}x = -\gamma(x - \theta)\mathrm{d}t + \sqrt{\kappa_2^2 x^2 + \kappa_1^2 x }\mathrm{d}W_t^{(2)}
\label{BPSDE}
\end{equation}
where $\mathrm{d}W_t^{(2)}$ is the normally distributed Wiener process, $\mathrm{d}W_t^{(2)} \sim \mathrm{N(}0,\, \mathrm{d}t \mathrm{)}$. Stock market researchers will immediately recognize that for $\kappa_2=0$, (\ref{BPSDE}) reduces to the Heston model of stochastic volatility \cite{heston1993closed, dragulescu2002probability} and for $\kappa_1=0$ to the multiplicative model of stochastic volatility \cite{nelson1990arch,fuentes2009universal,ma2014model}. The $\kappa_1=0$ model has also been used in the Bouchaud-M\'ezard network model of economic exchange \cite{bouchaud2000wealth, ma2013distribution}. We introduced the combined model (\ref{BPSDE}) \cite{dashti2018combined} in order to "marry" the Heston behavior, which seem to work better for long accumulations of stock returns, and the multiplicative behavior, which seems to work better for daily or a few-days returns.

The steady-state distribution of (\ref{BPSDE}) is a BP 
\begin{equation}
BP(x; p,q,\beta)=\frac{(1+\frac{x}{\beta})^{-p-q}(\frac{x}{\beta})^{-1+p}}{\beta B(p,q)}
\label{BPv}
\end{equation}
with the scale parameter
\begin{equation}
\label{betaBP}
\beta=\frac{\kappa_1^2}{\kappa_2^2}
\end{equation}
and the shape parameters
\begin{equation}
\label{pBP}
p=\frac{2 \gamma \theta}{\kappa_1^2}
\end{equation}
\begin{equation}
\label{qBP}
q=1+\frac{2 \gamma}{\kappa_2^2}
\end{equation}
In the context of wealth/income distribution, the first term in the r.h.s. of (\ref{BPSDE}) postulates the convergence to the mean value $\theta$ over the time scale $\propto \gamma^{-1}$ due to a simple give-and-take economic exchange. The divergence from the mean $\theta$ towards low end and high end are exclusively due to stochastic term, which is  $\propto x$ for large $x$ and are  $\propto \sqrt{x}$ for small $x$.  This term can be interpreted as fortunate and unfortunate events leading to gains and losses, such as inheritance, medical expenditures, bull and bear stock markets, etc. But they may also represent deviations from the mean due to work ethic, talent, etc. We underscore that eq. (\ref{BPSDE}) represents a time series for values $x$, which may not necessarily for a particular individual, but rather an abstract economic entity. The steady-state distributions represents, after the relaxation time, the distribution of values $x$ in this time series.

\subsection{Generalized Beta Prime SDE}

Consider now the following SDE
\begin{equation}
\mathrm{d}x = -\gamma(x - \theta x^{1-\alpha})\mathrm{d}t + \sqrt{\kappa_2^2 x^2 + \kappa_\alpha^2 x^ {2-\alpha}}\mathrm{d}W_t^{(2)}
\label{GB2SDE}
\end{equation}
Its steady-state distribution is \cite{hertzler2003classical, dashti2019implied}
\begin{equation}
GB2(x; p,q,\beta,\alpha)=\frac{\alpha (1+({\frac{x}{\beta}})^{\alpha})^{-p-q}(\frac{x}{\beta})^{-1+p\alpha}}{\beta B(p,q)}
\label{GB2x}
\end{equation}
with the scale parameter
\begin{equation}
\label{betaGB2}
\beta=(\frac{\kappa_\alpha}{\kappa_2})^{2 / \alpha}
\end{equation}
and shape parameters 
\begin{equation}
\label{pGB2}
p=\frac{1}{\alpha}(-1+\alpha +\frac{2 \gamma \theta}{\kappa_\alpha^2})
\end{equation}
and
\begin{equation}
\label{qGB2}
q=\frac{1}{\alpha}(1+\frac{2 \gamma}{\kappa_2^2})
\end{equation}
The steady-state distribution of (\ref{GB2SDE}) is GIGa for $\kappa_{\alpha}=0$ and GGa for $\kappa_2=0$. For $\alpha=1$ we have mean-reverting models which yield a BP steady-state distribution in general and IGa and Ga for $\kappa_1=0$ and $\kappa_2=0$ respectively. 

To understand the role of parameter $\alpha$, we analyze limiting behaviors (\ref{GB2largex}) and (\ref{GB2smallx}) vis-a-vis (\ref{GB2SDE}) using mean-reverting concept, even for $\alpha \neq 1$ and no fixed mean. Towards this end, we notice that from (\ref{qGB2}) the tail exponent $q \alpha$ does not depend on $\alpha$ so the high end is not affected, according to (\ref{GB2largex}). On the other hand, front exponent $p \alpha$ grows with alpha, so on transition from $\alpha < 1$ to $\alpha > 1$ the lower end gets suppressed and mid-range gets more populated, according to (\ref{GB2smallx}). This can be understood as follows: for $x \ll \beta$ in going from $\alpha < 1$ to $\alpha > 1$ we see the ``reversion" to the value smaller than the ``average" to larger than ``average" per $ \theta x^{1-\alpha}$ in (\ref{GB2SDE}), while volatility, including potential "losses," $\propto x^{2-\alpha}$, becomes smaller.\\

\section{Numerical Simulations \label{Simulations}}

We now undertake a numerical analysis of wealth/income distribution based on the sale prices of homes in Hamilton County, Ohio, which includes Cincinnati metropolitan area of about 2.2 million people. The data is for 1970 - 2010 sales and altogether we had 124,203 data points. We adjusted for inflation with the Bureau of Labor Statistics (BLS) Consumer Price Index (CPI) data for Hamilton County, expressing prices alternatively in 1990 and 2010 constant dollars to find only very minor differences -- see below.

\begin{figure}[!htbp]
\centering
\begin{tabular}{cc}
\includegraphics[width = 0.77 \textwidth]{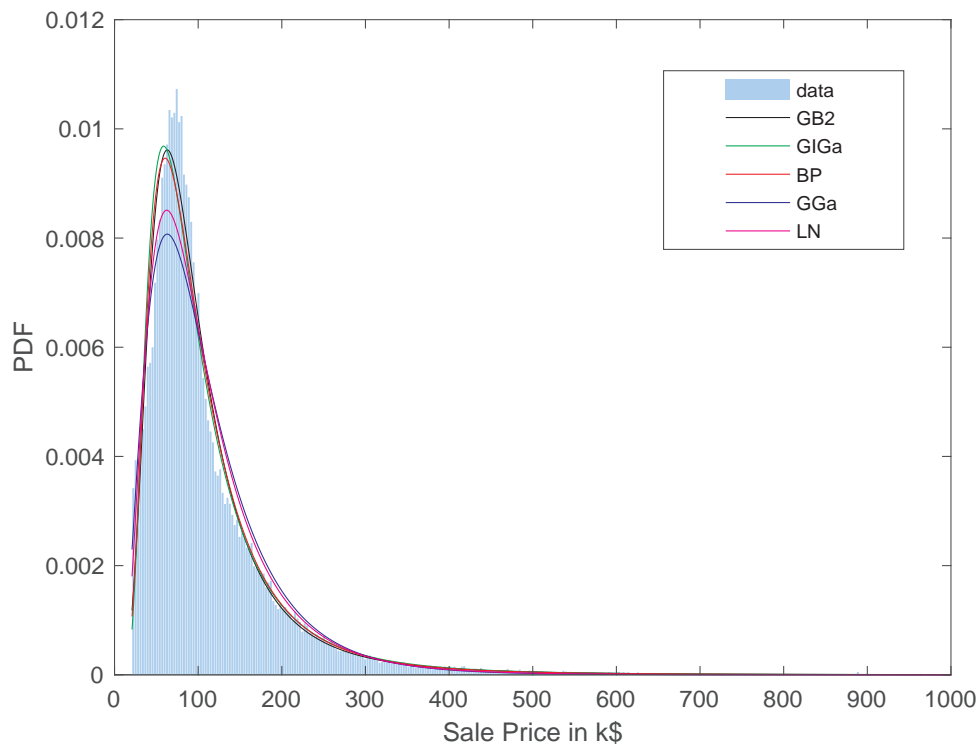} 
\end{tabular}
\caption{Fits of the sale price data in 1990 constant dollars.}
\label{Saleprice1990}
\end{figure}

\begin{figure}[!htbp]
\centering
\begin{tabular}{cc}
\includegraphics[width = 0.77 \textwidth]{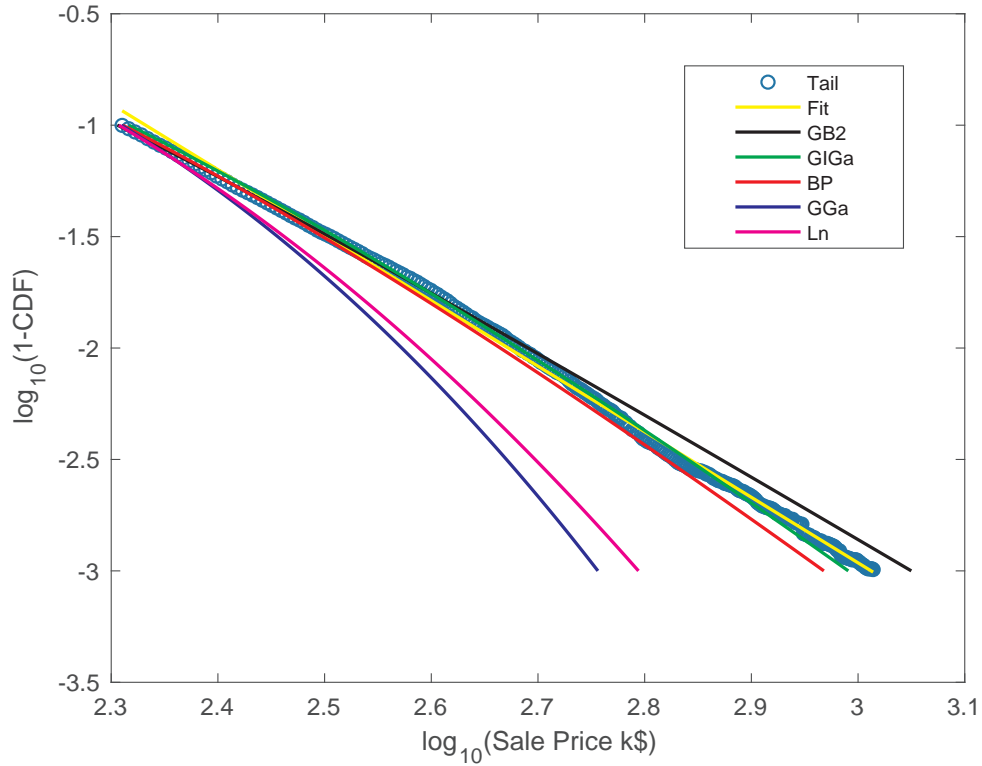}
\end{tabular}
\caption{Tail fitting using CDF}
\label{SalepriceTail1990}
\end{figure}

We use Maximum Likelihood Estimation (MLE) for fitting and the results of fitting are presented in Fig. \ref{Saleprice1990}, based on 1990 constant dollars, and Tables \ref{MLEsalepricedeinf1990} and \ref{MLEsalepricedeinf2010}, based on 1990 and 2010 constant dollars respectively. Additionally, we fit the tails of the Cumulative Density Function (CDF) -- assuming power-law tails -- and present the results in Fig. \ref{SalepriceTail1990}, based on 1990 constant dollars, and Tables \ref{SlopesalepricedeinfTail1990} and \ref{SlopesalepricedeinfTail2010}. For the actual fat-tailed distributions -- GB2, BP and GIGa -- the results of tail fits are compared with the tail parameters calculated from full distribution fits per Tables \ref{MLEsalepricedeinf1990} and \ref{MLEsalepricedeinf2010}. In our simulations we always computed using both 1990 and 2010 constant dollars to verify the consistency of our fitting, but Tables \ref{MLEsalepricedeinf1990} and \ref{MLEsalepricedeinf2010} amply demonstrate that, with the obvious exception of quantities related to the scale parameter, there are only very minor differences in quantities that depend on shape parameters. For this reason, in Figs. \ref{Saleprice1990} and \ref{SalepriceTail1990} and in what follows we present only1990 adjusted data. \\

\begin{table}[!htbp]
\centering
\caption{MLE results in 1990 constant dollars}
\label{MLEsalepricedeinf1990}
\resizebox{\textwidth}{!}{\begin{tabular}{ccccccccccc} 
\hline
            type &       parameters &          KS test &             Mean &              RMS &             Gini &   Hoover      & Theil T &Theil L &    DMMS\\
\hline
Data & N.A.& N.A. &         113.9110 &         132.2079 &           0.3620 &    0.2606      & 0.2580 &           0.2176 &    0.2708\\
GB2 & GB2(3.0300 , 1.5521 , 1.8265 , 57.5208) & 0.0173 & 113.7045 & 108.3528 & 0.3609 & 0.2580 & 0.2473 & 0.2168 &    0.1929\\
BP & BP(         13.3205,           3.7632,          23.4072) &           0.0228 &         112.8402 &          93.3789 &           0.3560 &    0.2556    &   0.2288 &           0.2082 &   0.1870  \\
GIGa & GIGa(          5.4618,         849.6230,           0.7200) &           0.0291 &         113.5776 &          97.3008 &           0.3624 &   0.2610      &  0.2393 &           0.2147 &   0.1999\\
LN & LN(          4.5178,           0.6200) &           0.0463 &         111.0540 &          76.0374 &           0.3389 &    0.2434    &    0.1922 &           0.1922 &   0.1551\\
GGa & GGa(         41.9754,           0.00001,           0.2442) &           0.0559 &         111.1247 &          74.1637 &           0.3390 &   0.2435     &   0.1897 &           0.1958 &   0.1353 \\
\hline
\end{tabular}}
\end{table}

\begin{table}[!htbp]
\centering
\caption{MLE results in 2010 constant dollars}
\label{MLEsalepricedeinf2010}
\resizebox{\textwidth}{!}{\begin{tabular}{cccccccccc} 
\hline
            type &       parameters &          KS test &             Mean &              RMS &             Gini &   Hoover   &      Theil T &          Theil L   &  DMMS \\
\hline
Data & N.A.& N.A. &         184.3548 &         213.9667 &           0.3620 &         0.2606 & 0.2580 &           0.2176 &      0.2639\\
GB2 & GB2(3.0302 , 1.5521 , 1.8265 , 93.0900) & 0.0173 & 184.0205 & 175.3528 & 0.3609 &   0.2580    &  0.2473 & 0.2168 &  0.1972\\
BP & BP(         13.3206,           3.7632,          37.8823) &           0.0228 &         182.6217 &          151.1257 &           0.3560 &   0.2556    &         0.2288 &           0.2082 &        0.1967 \\
GIGa & GIGa(          5.4628,        1375.4542,           0.7200) &           0.0291 &         183.8088 &         157.4365 &           0.3624 &    0.2610     &     0.2392 &           0.2146 &       0.1991\\
LN & LN(          4.9992,           0.6200) &           0.0463 &         179.7310 &         123.0598 &           0.3389 &    0.2435     &     0.1922 &           0.1922 &      0.1474\\
GGa & GGa(         44.6058,           0.00001,           0.2386) &           0.0546 &         179.2009 &         118.7474 &           0.3369 &   0.2434      &     0.1873 &           0.1930 &     0.1409\\
\hline
\end{tabular}}
\end{table}

\begin{table}[!htb]
\caption{Slope fitting results in 1990 constant dollars (left) and 2010 constant dollars (right).}
\label{MLESRV2OverVIX21990nn}
\begin{minipage}{0.5\textwidth}
\centering
\caption{Tail slope results in 1990 constant dollars}
\label{SlopesalepricedeinfTail1990}
\begin{tabular}{cccc} 
\hline
            type &            Slope &    param  & Slope  \\
\hline
Data &          -2.9411 &      & \\
GB2  &          -2.8253& $ -\alpha q$    & $-2.835$ \\
BP  &          -3.1102 &   $-q $ & $-3.763$\\
GIGa &          -3.0337  & $-\alpha \gamma $  & $-3.933$ \\
LN &          -4.1994 &     & \\
GGa  &          -4.5707 &    &\\
\hline
\end{tabular}
\end{minipage}
\begin{minipage}{.5\textwidth}
\centering
\caption{Tail slope results in 2010 constant dollars}
\label{SlopesalepricedeinfTail2010}
\begin{tabular}{cccc} 
\hline
            type &            Slope&  param & Slope  \\
\hline
Data &          -2.9411 &   & \\
GB2  &          -2.8253 & $-q \alpha$    & $-2.835$ \\
BP  &          -3.1102 &  $-q$ & $-3.763$\\
GIGa &          -3.0341 & $-\alpha \gamma$  &  $-3.933$ \\
LN &          -4.1994 &    & \\
GGa  &          -4.5899 &   & \\
\hline
\end{tabular}
  \end{minipage}

\end{table}

\section{Conclusions}
We argue that the unique property of Generalized Beta Prime (and Beta Prime), that makes it suitable for describing wealth/income and other distributions in natural and life sciences, is that it can mimic various behaviors, such as exponential, both for small and large variable. It also has an important property that the distribution of the inverse variable is also Generalized Beta Prime. In other words, small and large values of variables can be interchanged. Under such transformation, Generalized Beta Prime cleanly enforces the requirement on parameters to preserve bell shape and existence of the variance. Most importantly, however, is that it is a steady-state distribution of a simple stochastic model which can describe many phenomena, including modeling economic exchange and market volatility.

We investigated measures of income inequality, Gini, Hoover and Theil coefficients, in the Generalized Beta Prime framework and derived the relationships (\ref{GiniGB2transform}) and (\ref{TheilGB2transform}) (as well as (\ref{GiniBPtransform}) and (\ref{TheilPBtransform}) and connected them to the aforementioned property (\ref{GB2transform}) (and (\ref{BPtransform})) of transformation to the inverse variable. We also derived parameters of Generalized Beta Prime from those of the stochastic model and thoroughly analyzed how those parameters affect the distribution and the inequality measures. Additionally, we derived a simple but very precise approximation (\ref{GiniBPapprox}) to the exact Beta Prime Gini. We also derived several new Gini, Hoover and Theil coefficients for distributions in Tables \ref{Analyticforms2} and \ref{Analyticforms3}.

We argued that Gini and Theil T, by definition, exaggerate the influence power law dependencies of the distributions both for low end (underweigh) and especially for high end fat tails (overweigh). For this reason, we believe that Hoover (Pietra, Schultz) and Theil L are more appropriate measures in such cases, especially as far as fat tails are concerned. We also introduced a new scale-independent measure of inequality, DMMS, which estimates, via a distribution-agnostic procedure, the fraction of the low and high end wealth/income. We used Beta Prime to illustrate this measure. In numerical simulations, Hoover was consistently lower than Gini and DMMS smaller than Hoover (for all fitted distributions in the latter case). Likewise, Theil L was consistently lower than Theil T both for data and for all fat-tailed distribution fits.

We used Hamilton County, Ohio home sale prices as a proxy to wealth/income distribution. Given the very large data set, we used Maximum Likelihood Estimation to fit with a number of distributions. While Generalized Beta Prime provided the best fit, the absolute accuracy was not particularly high. Additionally, we fit the tails directly to query the correspondence between those fits and power-law exponents of full-distribution fits. We examined the effect of a tiny fraction of the highest sale prices and showed that cutting those from the distribution noticeably reduce the inequality indices and "fatness" of tails. Finally, we fitted the distribution of market values (asking prices) and found that their tails are considerably "fatter" than those of sale prices -- so much so that the theoretical variance does not exist. We argued that this is because market values are not described by a model of economic exchange and as such are "unphysical."

\appendix

\section{New Measure of Inequality \label{New}} 

\begin{figure}[!htbp]
\centering
\begin{tabular}{cc}
\includegraphics[width = 0.47 \textwidth]{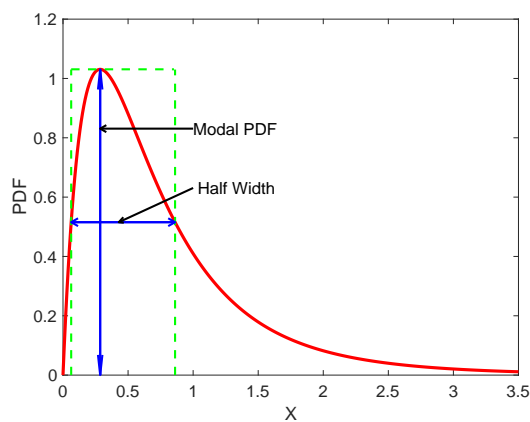}
\end{tabular}
\caption{Illustration of DMMS. Area inside represents roughly the proportion of wealth/income not in the tails.}
\label{Gini2graph}
\end{figure}

Here we introduce a new measure of inequality, DMMS, intended for bell-shaped distributions:
\begin{equation}
\label{Gini2}
DMMS=1- MPDF \times HW
\end{equation}
where $MPDF$ is the ``modal PDF", the height of the PDF at the mode, and $HW$ is the "half width", the width of the distribution at half modal PDF. This is illustrated in Fig. \ref{Gini2graph}, so that DMMS measures roughly the proportion of low and high end in the entire distribution. Since $MPDF \propto \beta^{-1}$ and $HW \propto \beta$, DMMS is scale-independent -- a requirement for inequality measure -- and is a number between 0 and 1.

\begin{figure}[!htbp]
\centering
\begin{tabular}{cc}
\includegraphics[width = 0.47 \textwidth]{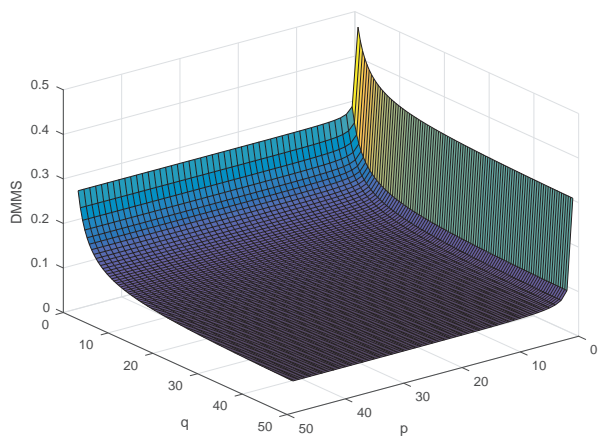}
\end{tabular}
\caption{Beta Prime DMMS as a function of $p$ and $q$.}
\label{Gini23D}
\end{figure}

We were unable to derive analytical expression for DMMS for BP (for GIGa, the closed-form expression is obtained in \cite{liu2016probability}), so we present only numerical results. With very high precision, the maximum value of DMMS is given by
\begin{equation}
\label{Gini2max}
D^{BP}_{max}=D^{BP} (1,2) \approx 0.48
\end{equation}
which is, as expected, smaller than maximum value of Gini and is slightly smaller than that of Hoover (\ref{GiniBPmax}). Fig. \ref{Gini23D} shows DMMS as a function of $p$ and $q$. It is similar to Gini and Hoover, but with lower values.

\section{Expanded Table \label{Expanded}}
In Tables \ref{Analyticforms4} and \ref{Analyticforms5}, we add IGa and Ga as limiting cases of GIGa and GGa respectively. BP, IGa and Ga are obtained from GB2, GIGa and GGa by setting one of the shape parameters to unity. Often times, fitting with reduced distributions is close in accuracy, while the expressions for the indices are greatly simplified relative to their generalized counterparts.

\begin{sidewaystable}
\centering
\caption{Analytic Forms of Distributions}
\label{Analyticforms4}
\fontsize{7.5}{8.0}
\begin{tabular}{|c|c|c|c|} 
\hline
            type &       PDF &        CDF &          Mean ($\mu$)  \\
\hline
GB2 & $\frac{\alpha(1+(\frac{x}{\beta})^{\alpha})^{-p-q}(\frac{x}{\beta})^{-1+p\alpha}}{\beta \space \Beta(p,q)}$ &$I(\frac{x^\alpha}{x^\alpha+\beta^\alpha},p,q)$ &  $\frac{\beta \Beta(p + \frac{1}{\alpha},q - \frac{1}{\alpha})}{\Beta(p,q)}$\\
\hline
BP & $ \frac{(1+\frac{x}{\beta})^{-p-q}(\frac{x}{\beta})^{-1+p}}{\beta \space B(p,q)}$ &  $I(\frac{x}{x+\beta},p,q)$&$\frac{\beta p}{q -1}$ \\
\hline
GIGa & $\frac{\gamma e^{-(\frac{\beta}{x})^\gamma}(\frac{\beta}{x})^{1+\alpha\gamma}}{\beta \Gamma(\alpha)}$& $Q(\alpha,(\frac{\beta}{x})^\gamma) $ &$\frac{\beta \Gamma(\alpha - \frac{1}{\gamma})}{\Gamma(\alpha)}$  \\
\hline
IGa &  $\frac{e^{-\frac{\beta}{x}}(\frac{\beta}{x})^{1+\alpha}}{\beta \Gamma(\alpha)}$& $Q(\alpha,\frac{\beta}{x}) $ &$\frac{\beta}{\alpha -1}$  \\
\hline
GGa & $\frac{\gamma e^{-(\frac{x}{\beta})^{\gamma}}(\frac{x}{\beta})^{-1+\alpha \gamma}}{ \beta \Gamma(\alpha)}$ & $1 - Q(\alpha,(\frac{x}{\beta})^\gamma)$ & $\frac{\beta \Gamma(\alpha + \frac{1}{\gamma})}{\Gamma(\alpha)}$  \\
\hline
Ga &  $\frac{e^{-\frac{x}{\beta}}(\frac{x}{\beta})^{-1+\alpha}}{ \beta \Gamma(\alpha)}$ & $1 - Q(\alpha,\frac{x}{\beta})$ & $\alpha \beta$  \\
\hline
\end{tabular}

    \vspace{2\baselineskip}
    \centering
\caption{Gini and Hoover Indices}
\label{Analyticforms5}
\fontsize{7.5}{8.0}
\begin{tabular}{|c|c|c|} 
\hline
            type &       Gini &        Hoover  \\
\hline
GB2 &  \makecell{$\frac{\Beta(2q-\frac{1}{\alpha},2p+\frac{1}{\alpha})}{\Beta(p,q) \Beta(p+\frac{1}{\alpha},q-\frac{1}{\alpha})} (\frac{1}{p} \prescript{}{3}{F}{}_{2}(1,p+q,2p+\frac{1}{\alpha};p+1,2(p+q);1)-$  \\$\frac{1}{p+\frac{1}{\alpha}}\prescript{}{3}{F}{}_{2}(1,p+q,2p+\frac{1}{\alpha};p+1+\frac{1}{\alpha},2(p+q);1))$} & $I(\frac{(\frac{\mu}{\beta})^{\alpha}}{1+(\frac{\mu}{\beta})^{\alpha}},p,q)- I(\frac{(\frac{\mu}{\beta})^{\alpha}}{1+(\frac{\mu}{\beta})^{\alpha}},p+\frac{1}{\alpha},q-\frac{1}{\alpha})$ \\
\hline
BP &  \makecell{$\frac{\Beta(2p+1,2q-1)}{\Beta(p,q) \Beta(p+1,q-1)} (\frac{1}{p} \prescript{}{3}{F}{}_{2}(1,p+q,2p+1;p+1,2(p+q);1)-$  \\$\frac{1}{p+1}\prescript{}{3}{F}{}_{2}(1,p+q,2p+1;p+2,2(p+q);1))$\\$=\frac{\Beta(2p + 1,2q-1)}{\Beta(p,q) \Beta(p+1,q-1)}\frac{2p+2q-1}{p(q-1)}$} &$I(\frac{\frac{\mu}{\beta}}{1+\frac{\mu}{\beta}},p,q) - I(\frac{\frac{\mu}{\beta}}{1+\frac{\mu}{\beta}},p+1,q-1) = \frac{p^{-1 + p} (-1 + q)^{-1 + q} (-1 + p + q)^{1 - p - q}}{\Beta(p,q)}$ \\
\hline
GIGa &  \makecell{$\frac{1}{\Beta(\alpha,\alpha-\frac{1}{\gamma})}(\frac{1}{\alpha-\frac{1}{\gamma}} \prescript{}{2}{F}{}_{1}(\alpha-\frac{1}{\gamma},2 \alpha-\frac{1}{\gamma};\alpha-\frac{1}{\gamma}+1;-1) - $\\$(\frac{1}{\alpha}) \prescript{}{2}{F}{}_{1}(\alpha,2 \alpha-\frac{1}{\gamma};\alpha + 1;-1))$} &  $Q(\alpha,(\frac{\beta}{\mu})^\gamma) - Q(\alpha- \frac{1}{\gamma},(\frac{\beta}{\mu})^\gamma)$\\
\hline
IGa &  \makecell{$\frac{1}{\Beta(\alpha,\alpha-1)}(\frac{1}{\alpha-1} \prescript{}{2}{F}{}_{1}(\alpha-1,2 \alpha-1;\alpha;-1) - $\\$(\frac{1}{\alpha}) \prescript{}{2}{F}{}_{1}(\alpha,2 \alpha-1;\alpha + 1;-1))$\\$=\frac{\Gamma(\alpha-\frac{1}{2})}{\sqrt{\pi}\Gamma(\alpha)}$} &  \makecell{$Q(\alpha,(\frac{\beta}{\mu})) - Q(\alpha- 1,(\frac{\beta}{\mu}))$\\$= \frac{e^{-\frac{\beta}{\mu}} (\frac{\beta}{\mu})^{\alpha-1}}{\Gamma(\alpha)}$}\\
\hline
GGa & \makecell{$\frac{1}{2^{2 \alpha + \frac{1}{\gamma}} \Beta(\alpha,\alpha+\frac{1}{\gamma})}(\frac{1}{\alpha} \prescript{}{2}{F}{}_{1}(1,2 \alpha+\frac{1}{\gamma};\alpha+1;\frac{1}{2}) - $\\$(\frac{1}{\alpha+\frac{1}{\gamma}}) \prescript{}{2}{F}{}_{1}(1,2 \alpha+\frac{1}{\gamma};\alpha + \frac{1}{\gamma};\frac{1}{2}))
$ } & $ Q(\alpha + \frac{1}{\gamma},(\frac{\mu}{\beta})^\gamma)- Q(\alpha,(\frac{\mu}{\beta})^\gamma)$\\
\hline
Ga & \makecell{$\frac{1}{2^{2 \alpha + 1} \Beta(\alpha,\alpha+1)}(\frac{1}{\alpha} \prescript{}{2}{F}{}_{1}(1,2 \alpha+1;\alpha+1;\frac{1}{2}) - $\\$(\frac{1}{\alpha+1}) \prescript{}{2}{F}{}_{1}(1,2 \alpha+1;\alpha + 1;\frac{1}{2}))
$ \\$=\frac{\Gamma(\alpha + \frac{1}{2})}{\sqrt{\pi}\Gamma(\alpha + 1)}$} & \makecell{$Q(\alpha + 1,(\frac{\mu}{\beta}))- Q(\alpha,(\frac{\mu}{\beta}))$\\$= \frac{e^{-\frac{\mu}{\beta}} (\frac{\mu}{\beta})^\alpha}{\Gamma(\alpha + 1)}$}\\
\hline
\end{tabular}

\begin{tablenotes}
  \item[*] $B(p,q) = \frac{\Gamma(p)\Gamma(q)}{\Gamma(p+q)}$: beta function; $\Gamma(\alpha)$: gamma function.
  \item[a] $\psi(x) = \frac{\mathrm{d}\ln\Gamma(x)}{\mathrm{d}x} = \frac{\Gamma'(x)}{\Gamma(x)}$: digamma function.
  \item[d] $Q(\alpha,x)= \frac{\Gamma(\alpha,x)}{\Gamma(\alpha)}$: regularized gamma function; $\Gamma(\alpha,x)$: incomplete gamma function
   \item[e] $I(x,p,q)=\frac{\Beta(x,p,q)}{\Beta(p,q)}$: regularized beta function; $\Beta(x,p,q)$: incomplete beta function
  \end{tablenotes}
   \end{sidewaystable}

\section{Tail Cuts \label{Tailcuts}}
\begin{figure}[!htbp]
\centering
\begin{tabular}{cc}
\includegraphics[width = 0.49 \textwidth]{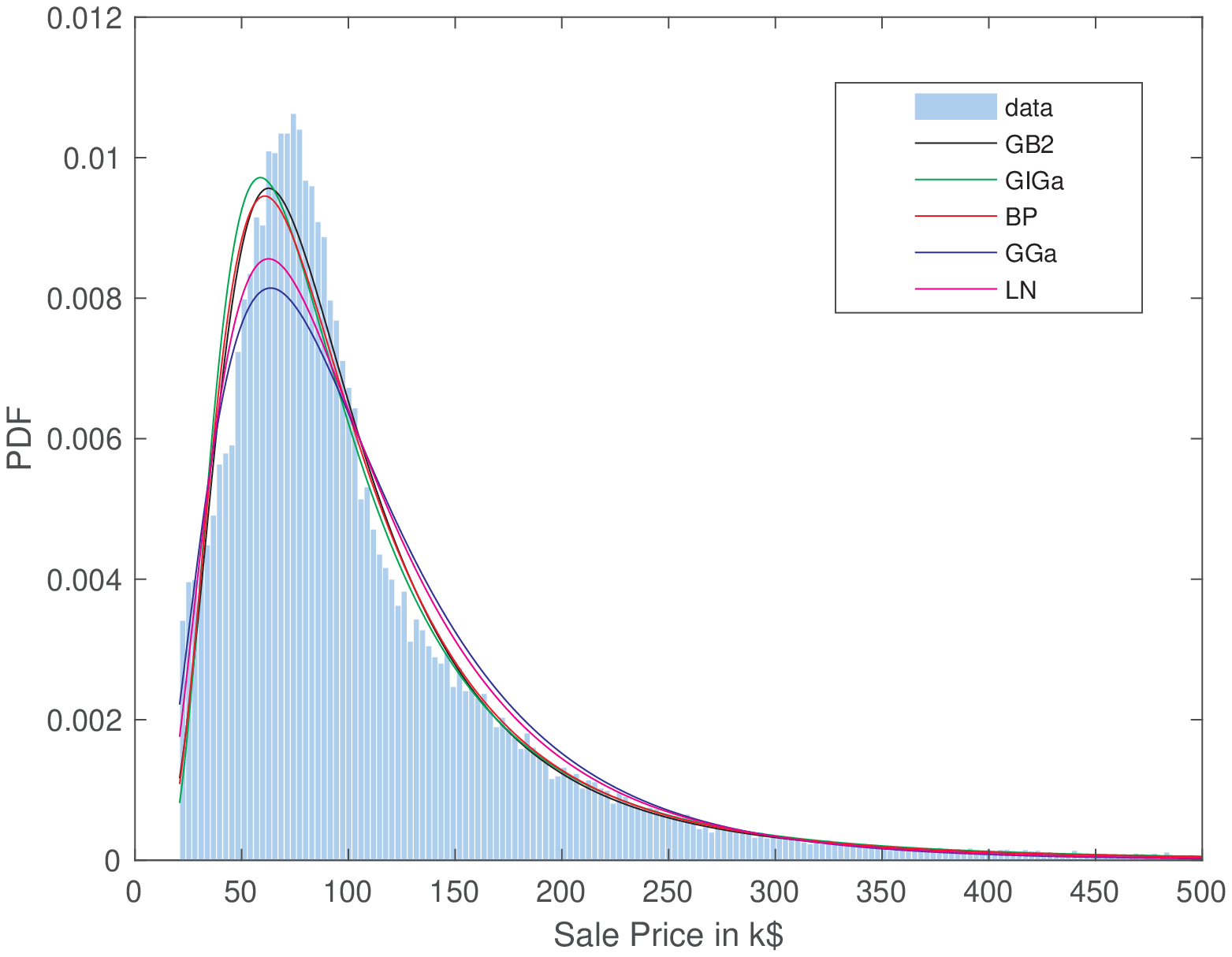}
\includegraphics[width = 0.49 \textwidth]{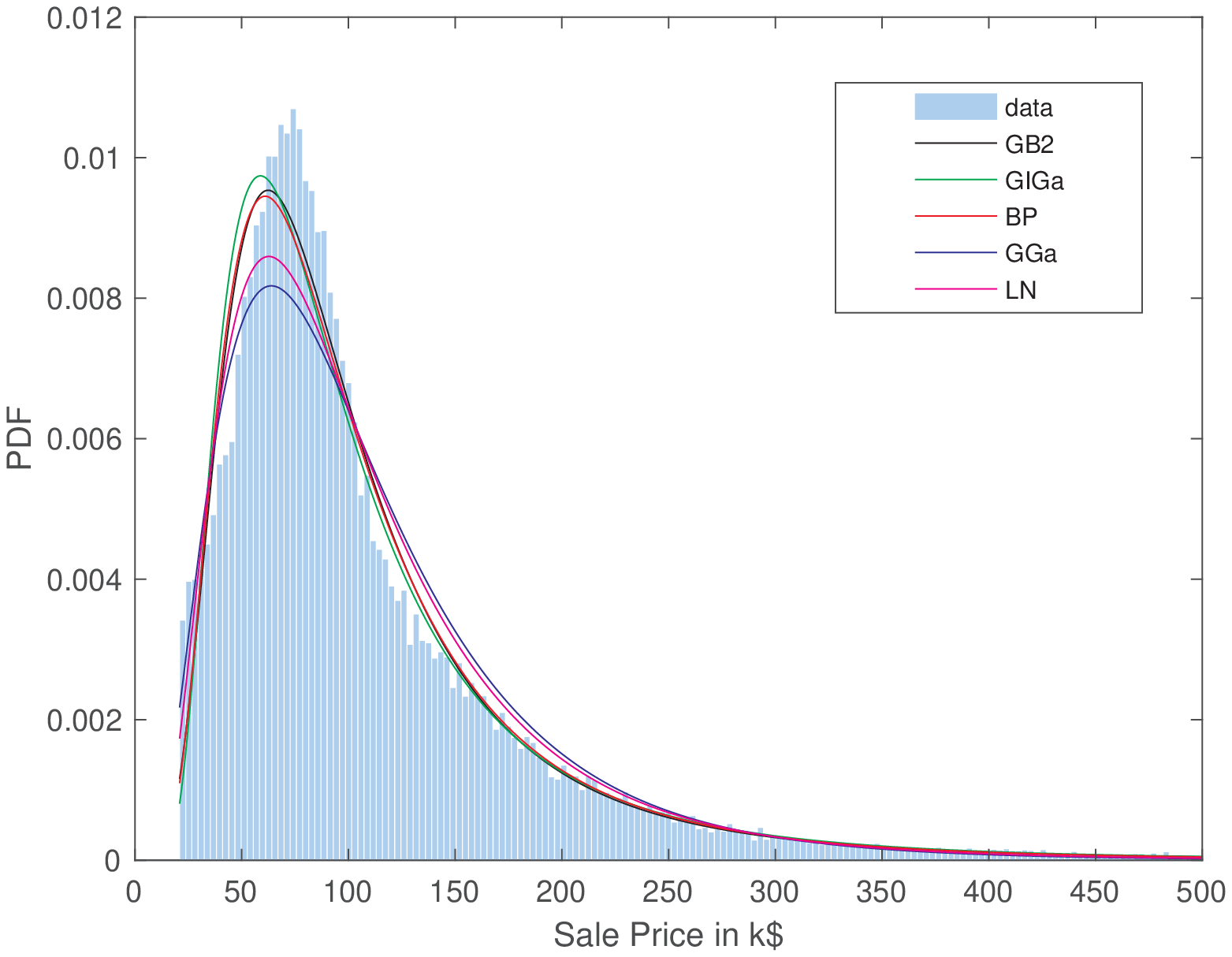}
\end{tabular}
\caption{MLE fits with the top 0.05\% sale prices cut (left) and top 0.1\% cut (right.)}
\label{Saleprice199005}
\end{figure}

\begin{figure}[!htbp]
\centering
\begin{tabular}{cc}
\includegraphics[width = 0.49 \textwidth]{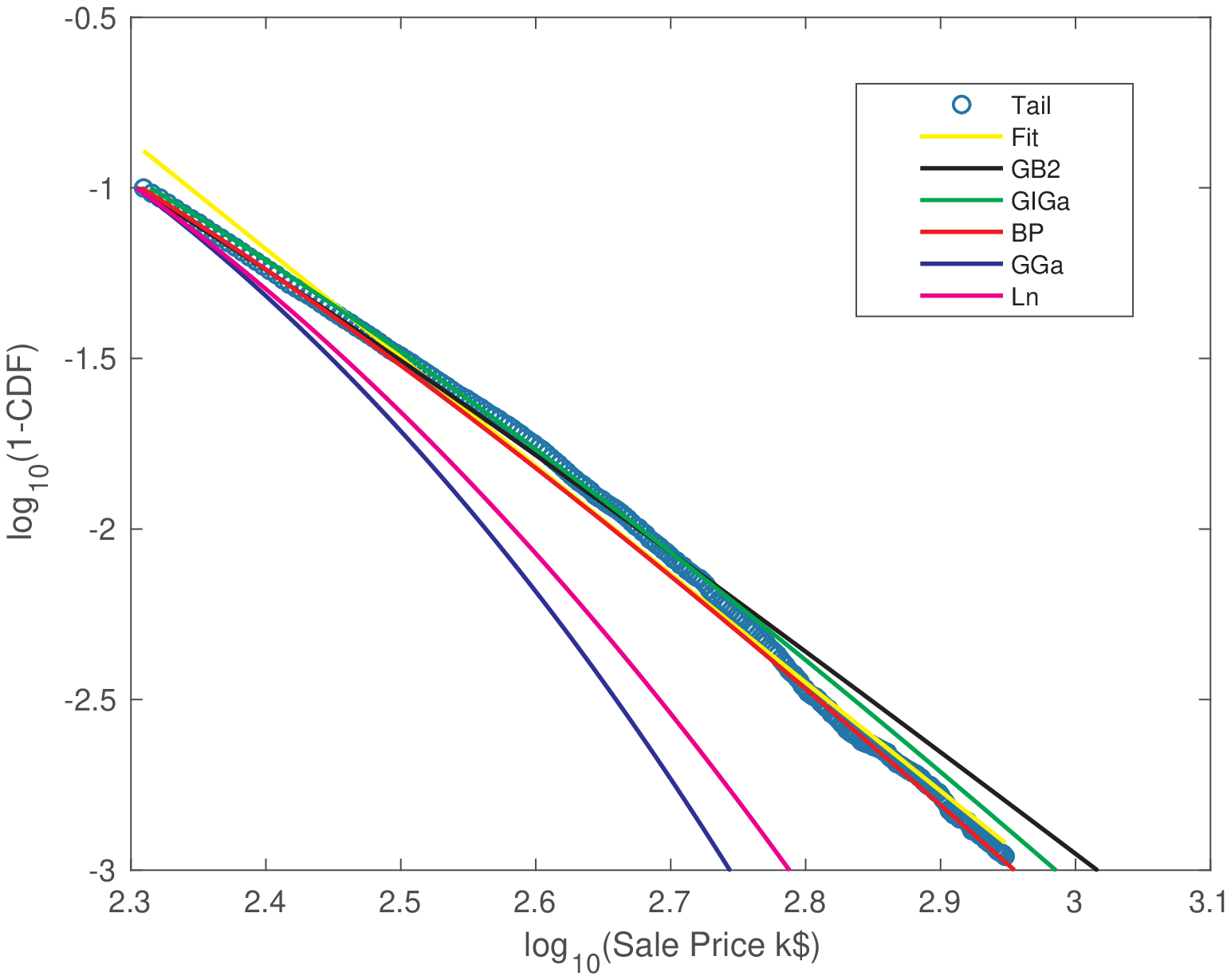}
\includegraphics[width = 0.49 \textwidth]{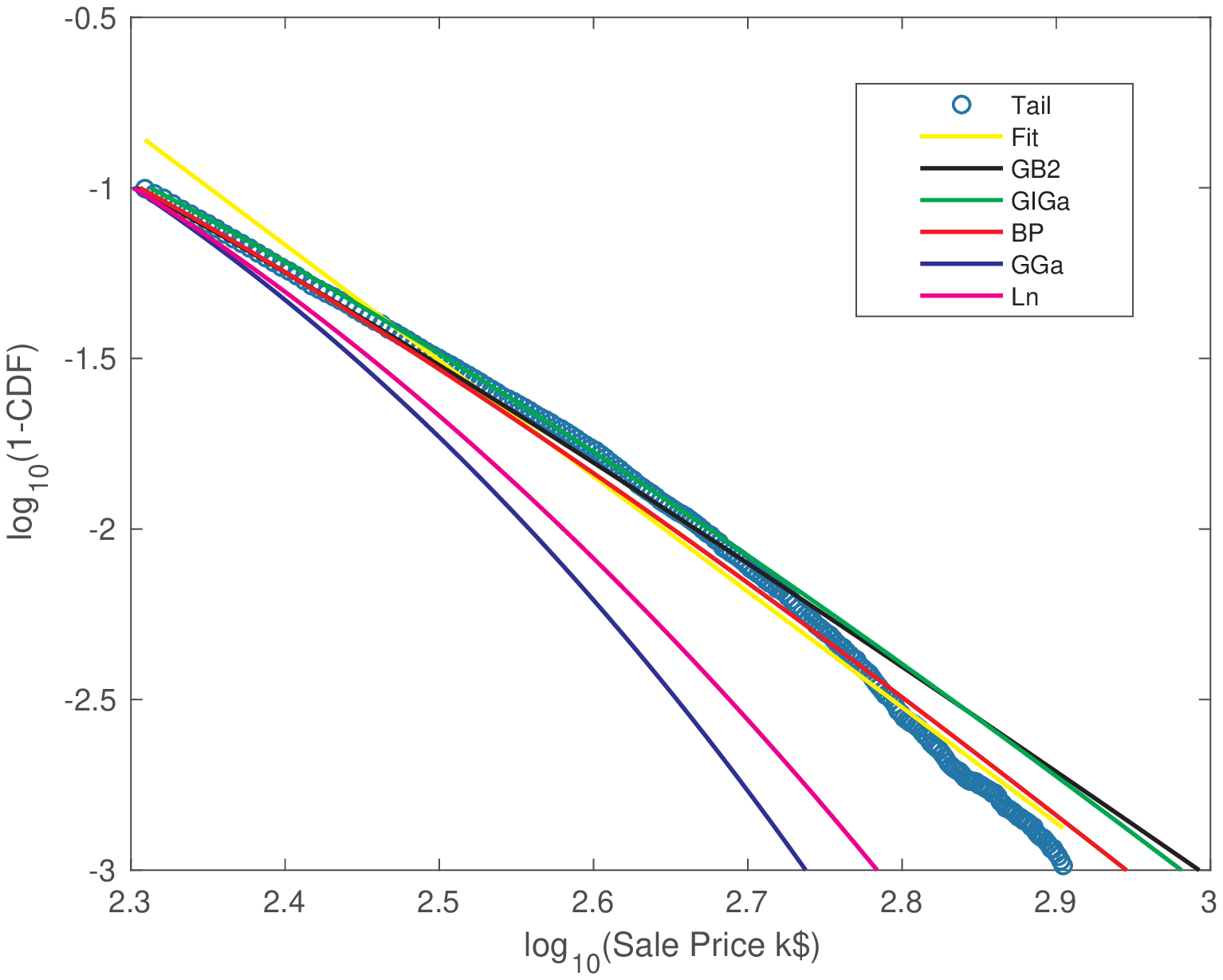}
\end{tabular}
\caption{Tail slope fits with the top 0.05\% sale prices cut (left) and top 0.1\% cut (right.)}
\label{SalepriceTail1990005}
\end{figure}

Here we investigate to what extent the top sale prices affect the distributions, their tails and inequality measures. Towards that end, we cut off first the top 0.05\% and then 0.1\% of sale prices. The results are presented in Figs. \ref{Saleprice199005} and \ref{SalepriceTail1990005} and Tables \ref{MLEsalepricedeinf199005}, \ref{MLEsalepricedeinf19901} and \ref{SlopesalepricedeinfTail199005}. Remarkably, the first cut noticeably lowers inequality indices and makes tails less "fat." The second cut does not have as much of an effect relative to the first.\\

\begin{table}[!htbp]
\centering
\caption{MLE results for sales prices with top 0.05\% cut}
\label{MLEsalepricedeinf199005}
\resizebox{\textwidth}{!}{\begin{tabular}{ccccccccc} 
\hline
            type &       parameters &          KS test &             Mean &              RMS &             Gini &   Hoover      & Theil T &          Theil L  \\
\hline
Data & N.A.& N.A. &         112.4606 &          89.6130 &           0.3541 &     0.2554       & 0.2255 &           0.2064\\
GB2 & GB2(          3.8260,           1.8778,           1.6191,          53.6299) &           0.0176 &         112.9118 &         100.8824 & 0.3567 &    0.2552     &  0.2179 &           0.2302\\
GIGa & GIGa(          5.5018,         857.7555,           0.7200) &           0.0290 &         113.1787 &          96.2318 &           0.3609 &   0.2599       & 0.2369 &           0.2128\\
BP & BP(         12.5840,           3.8592,          25.5170) &           0.0229 &         112.3070 &          91.2442 &           0.3532 &      0.2535    & 0.2242 &           0.2051  \\
LN & LN(          4.5162,           0.6159) &           0.0452 &         110.5885 &          75.1049 &           0.3368 &   0.2419      &  0.1896 &           0.1896 \\
GGa & GGa(         39.3820,           0.0001,           0.2553) &           0.0543 &         110.3299 &          72.4940 &           0.3350 &    0.2405      & 0.1849 &           0.1909 \\
\hline
\end{tabular}}
\end{table}

\begin{table}[!htbp]
\centering
\caption{MLE results for sales prices with top 0.1\% cut}
\label{MLEsalepricedeinf19901}
\resizebox{\textwidth}{!}{\begin{tabular}{ccccccccc} 
\hline
            type &       parameters &          KS test &             Mean &              RMS &             Gini &   Hoover    &   Theil T &          Theil L  \\
\hline
Data & N.A.& N.A. &         111.9174 &          86.4710 &           0.3514 &         0.2535 & 0.2191 &           0.2029\\
GB2 & GB2(          4.5652,           2.1658,           1.4828,          50.4532) &           0.0163&         112.3574 &          96.5217 & 0.3539 & 0.2533       &   0.2289 &           0.2087\\
GIGa & GIGa(          5.5312,         863.5680,           0.7200) &           0.0290 &         112.8726 &          95.4512 &           0.3597 &        0.2590 &  0.2352 &           0.2114\\
BP & BP(         12.1545,           3.9269,          26.9504) &           0.0229 &         111.9161 &          89.8078 &           0.3514 &    0.2521       &  0.2210 &           0.2029  \\
LN & LN(          4.5148,           0.6133) &           0.0444 &         110.2659 &          74.5097 &           0.3355 &  0.2409      &   0.1881 &           0.1881 \\
GGa & GGa(         37.8721,           0.0001,           0.2621) &           0.0540 &         109.9910 &          71.6878 &           0.3329 &     0.2390    &    0.1825 &           0.1885 \\
\hline
\end{tabular}}
\end{table}

\clearpage

\begin{table}[!htb]
\caption{Tail slope results  for sales prices with top 0.05\% cut (left) and 0.1\% cut (right)}
\label{SlopesalepricedeinfTail199005}
\begin{minipage}{0.5\textwidth}
\begin{center}
\begin{tabular}{ c c} 
\multicolumn{2}{c}{} \\
\hline
            type &            Slope  \\
\hline
Data &          -3.1776 \\
GB2  &          -2.8573 \\
BP  &          -3.1613 \\
GIGa &          -3.0494 \\
LN &          -4.2272 \\
GGa  &          -4.6416 \\
\hline
\end{tabular}
\end{center}
\end{minipage}
\begin{minipage}{.5\textwidth}
\begin{center}
\begin{tabular}{ c c} 
\multicolumn{2}{c}{} \\
\hline
            type &            Slope  \\
\hline
Data &          -3.3890 \\
GB2  &          -2.9539 \\
BP  &          -3.1968 \\
GIGa &          -3.0609 \\
LN &          -4.2446 \\
GGa  &          -4.6801 \\
\hline
\end{tabular}
\end{center}
  \end{minipage}
\end{table}

\clearpage

\section{Tails of Top Properties}
Here we do tail fitting of the top 20\% and 30\% sale prices. We do not find significant differences with Tables \ref{SlopesalepricedeinfTail1990} and \ref{SlopesalepricedeinfTail2010}.\\

\begin{figure}[!htbp]
\centering
\begin{tabular}{cc}
\includegraphics[width = 0.49 \textwidth]{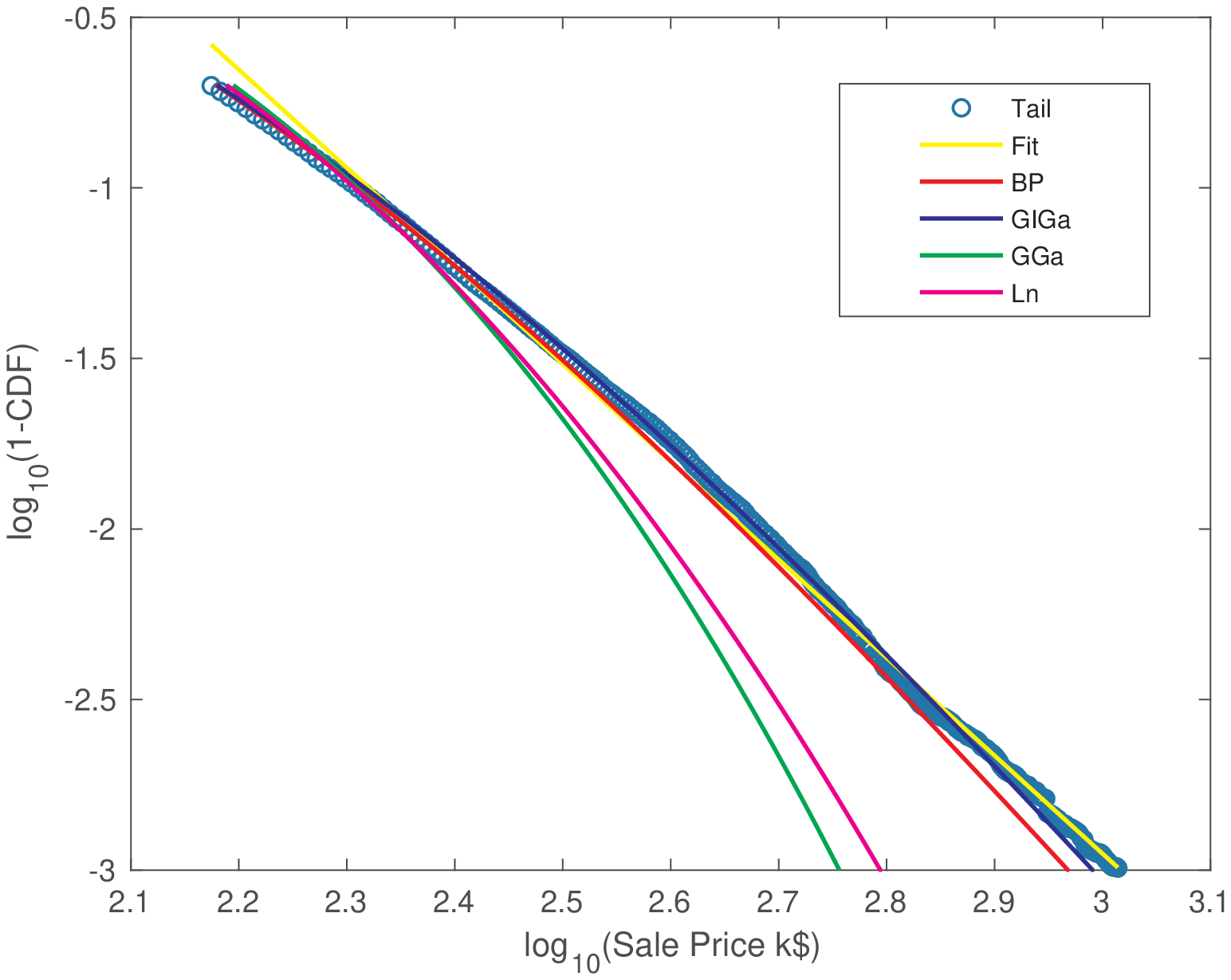}
\includegraphics[width = 0.49 \textwidth]{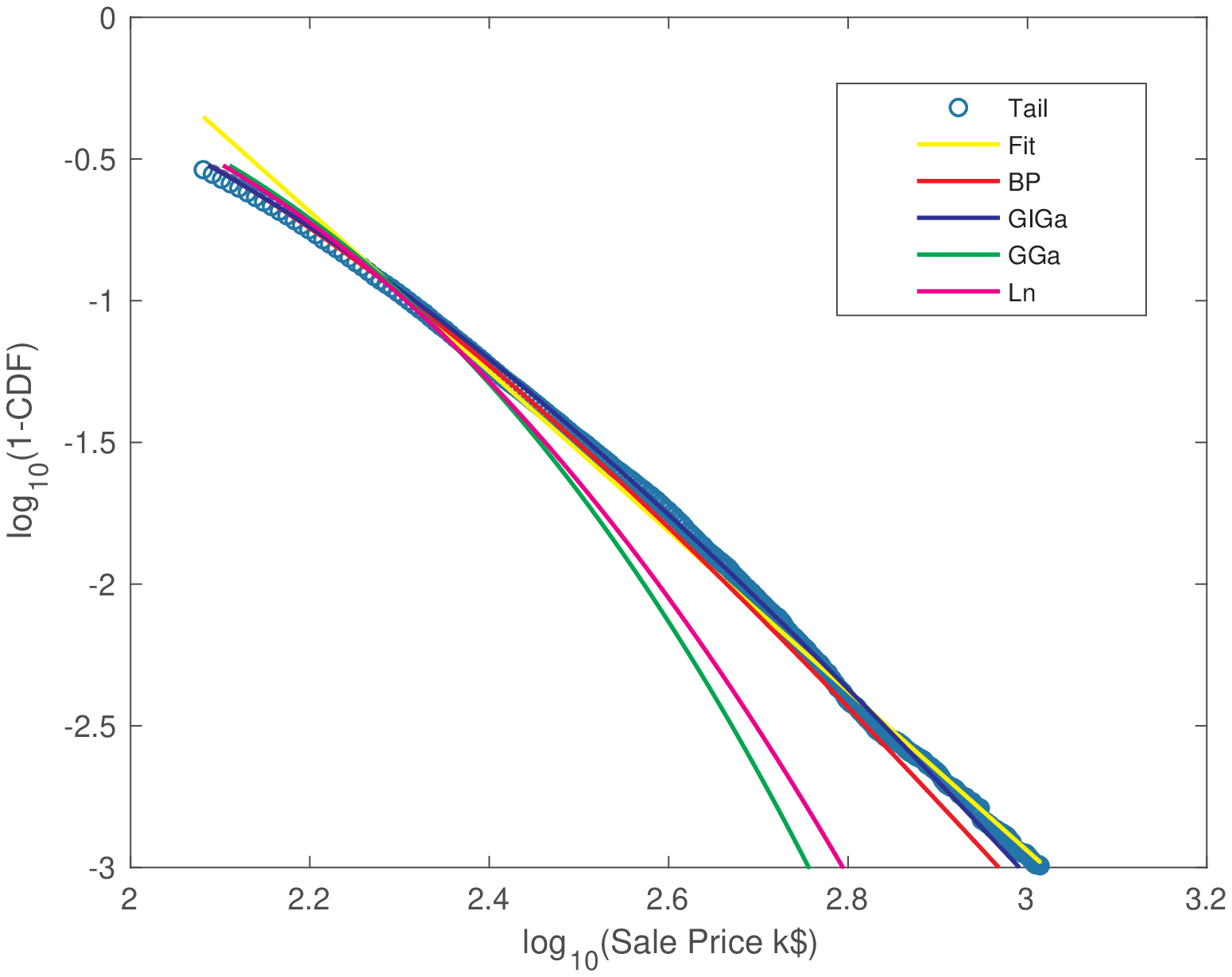}
\end{tabular}
\caption{Tail slope fits for sales prices of top 20\% prices (left) and 30\% cut (right).}
\label{Saleprice1990Tail}
\end{figure}

\begin{table}[!htb]
\caption{Tail slope results for sales prices of top 20\% prices (left) and 30\% cut (right)}
\label{SlopesalepricedeinfTail19901}
\begin{minipage}{0.5\textwidth}
\begin{center}
\begin{tabular}{ c c} 
\multicolumn{2}{c}{} \\
\hline
            type &            Slope  \\
\hline
Data &          -2.8730 \\
BP  &          -3.0229 \\
GIGa &          -2.9451 \\
LN &          -3.9405 \\
GGa  &          -4.2379 \\
\hline
\end{tabular}
\end{center}
\end{minipage}
\begin{minipage}{.5\textwidth}
\begin{center}
\begin{tabular}{ c c} 
\multicolumn{2}{c}{} \\
\hline
            type &            Slope  \\
\hline
Data &          -2.8189 \\
BP  &          -2.9594 \\
GIGa &          -2.8822 \\
LN &          -3.7644 \\
GGa  &          -4.0129 \\
\hline
\end{tabular}
\end{center}
  \end{minipage}

\end{table}






\clearpage
\section{Market Value Distribution \label{Market}}

\begin{figure}[!htbp]
\centering
\begin{tabular}{cc}
\includegraphics[width = 0.7 \textwidth]{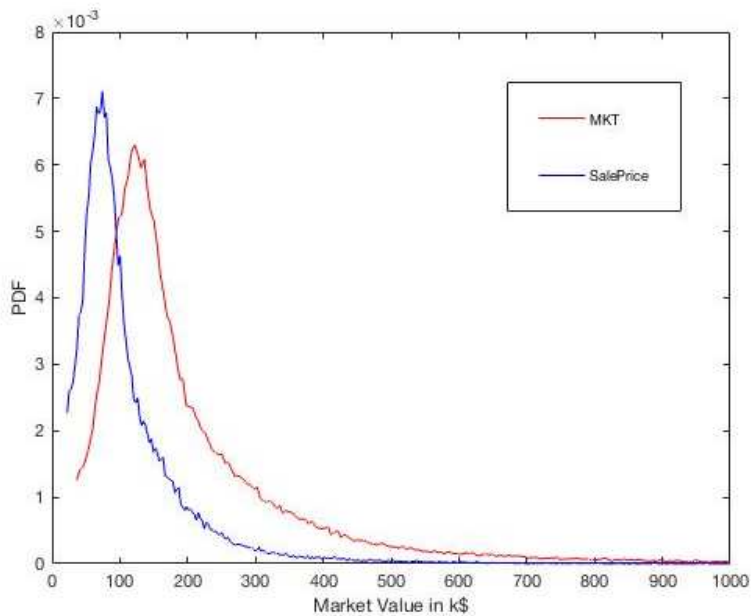}
\end{tabular}
\caption{Contour plots of the distributions of sale prices and market values.}
\label{MKTSP}
\end{figure}

\begin{figure}[!htbp]
\centering
\begin{tabular}{cc}
\includegraphics[width = 0.49 \textwidth]{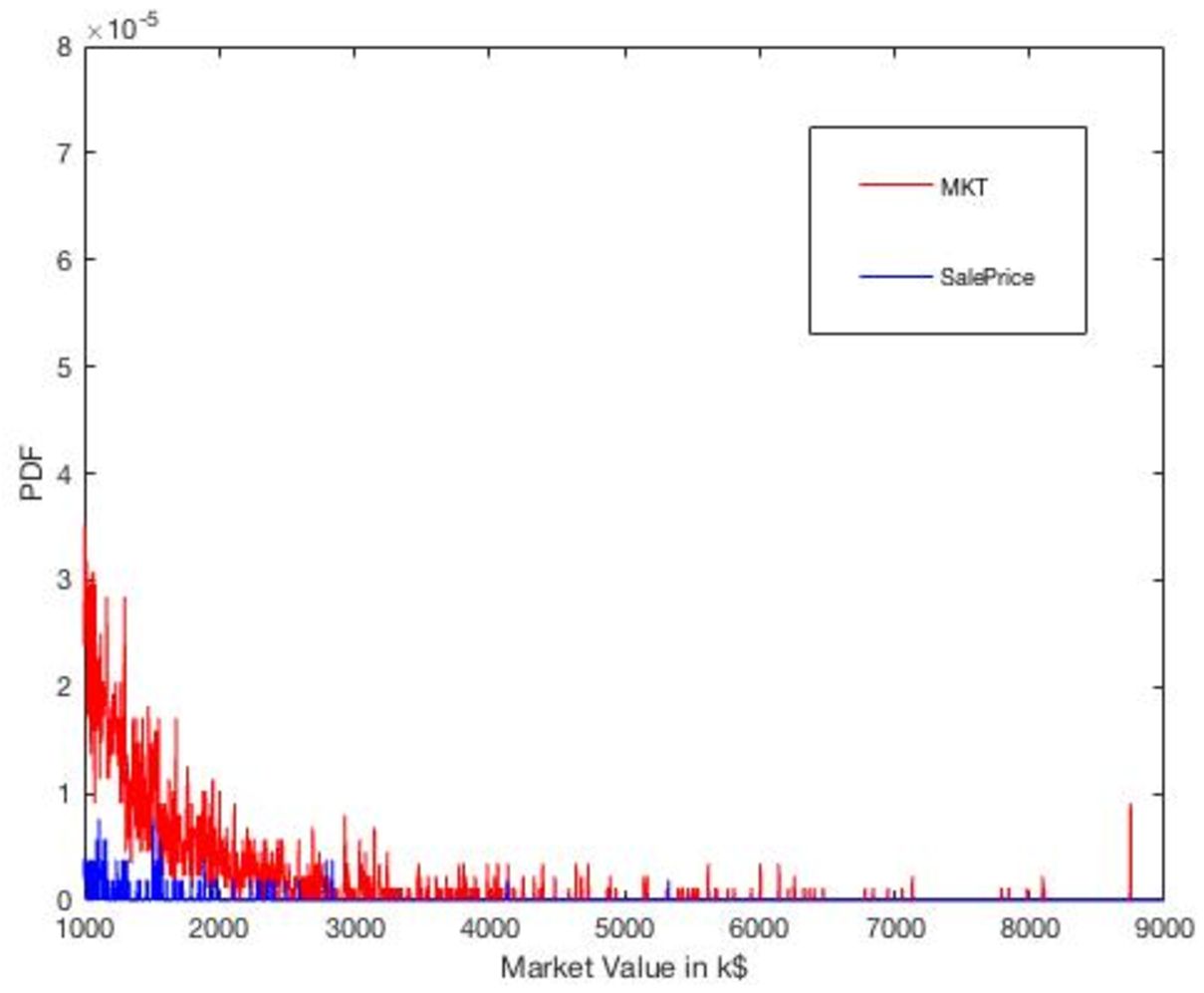}
\includegraphics[width = 0.49 \textwidth]{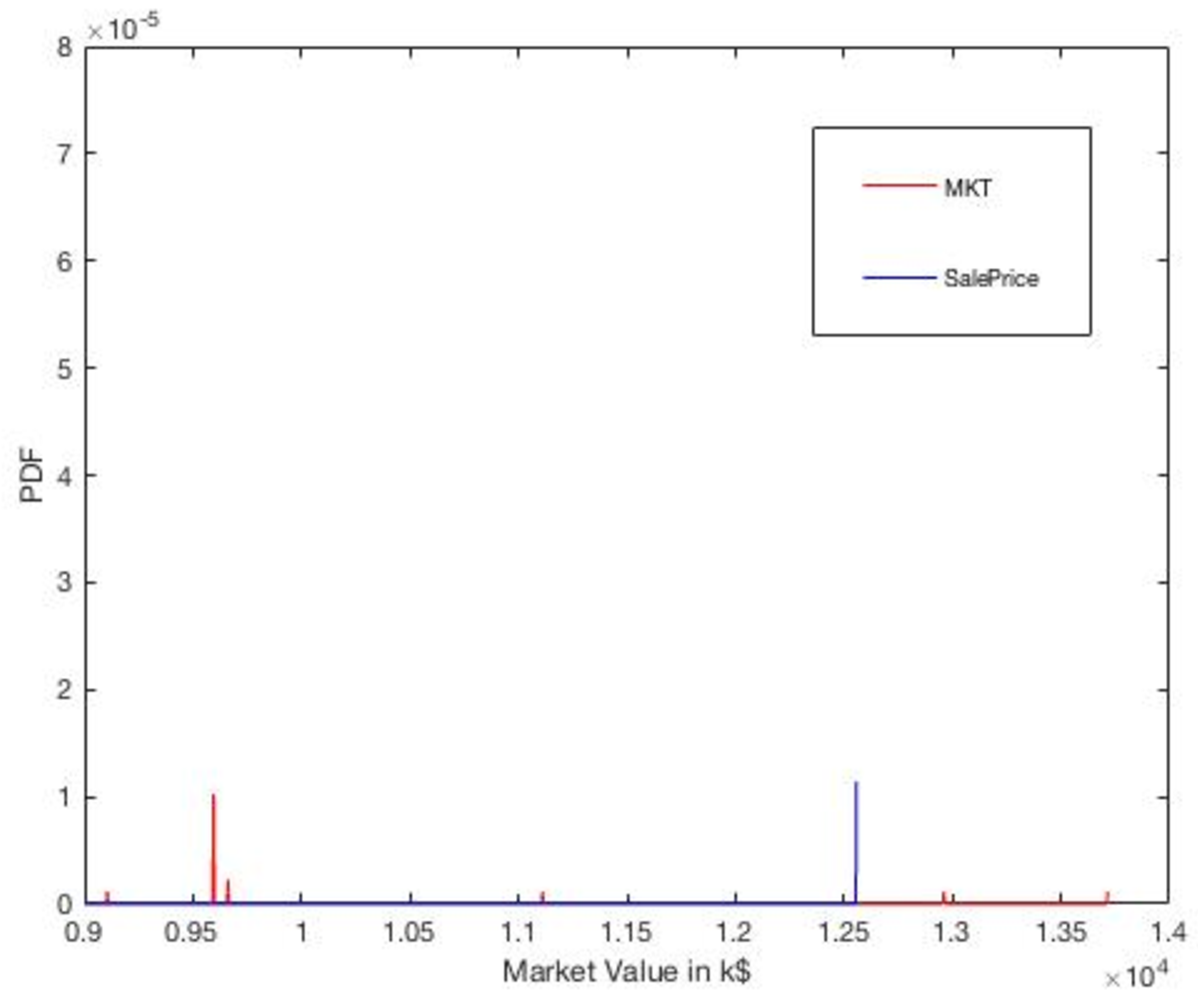}
\end{tabular}
\caption{Same as Fig. \ref{MKTSP} for high prices (left) and top prices (right).}
\label{MKTSPtails}
\end{figure}

\begin{figure}[!htbp]
\centering
\begin{tabular}{cc}
\includegraphics[width = 0.49 \textwidth]{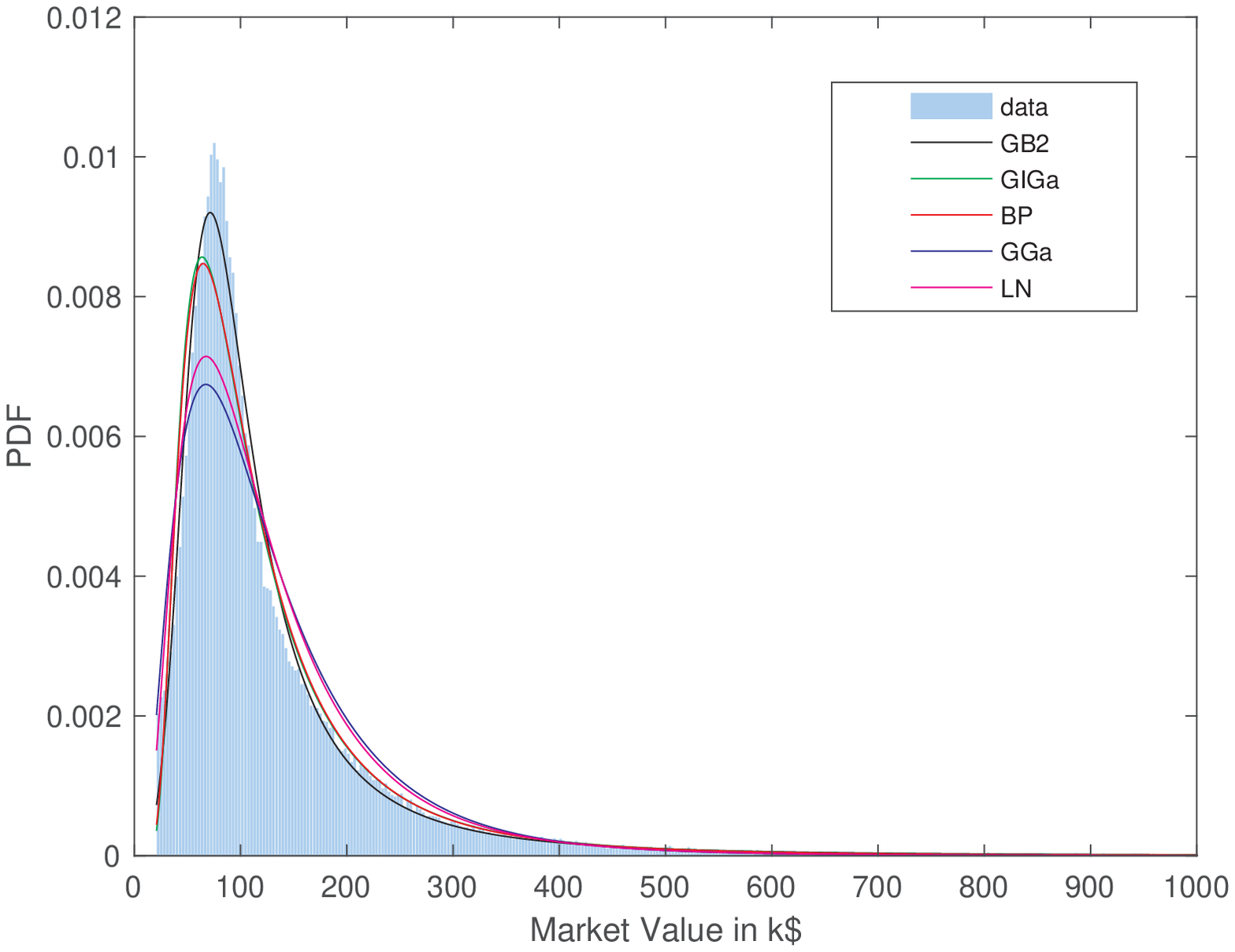}
\includegraphics[width = 0.49 \textwidth]{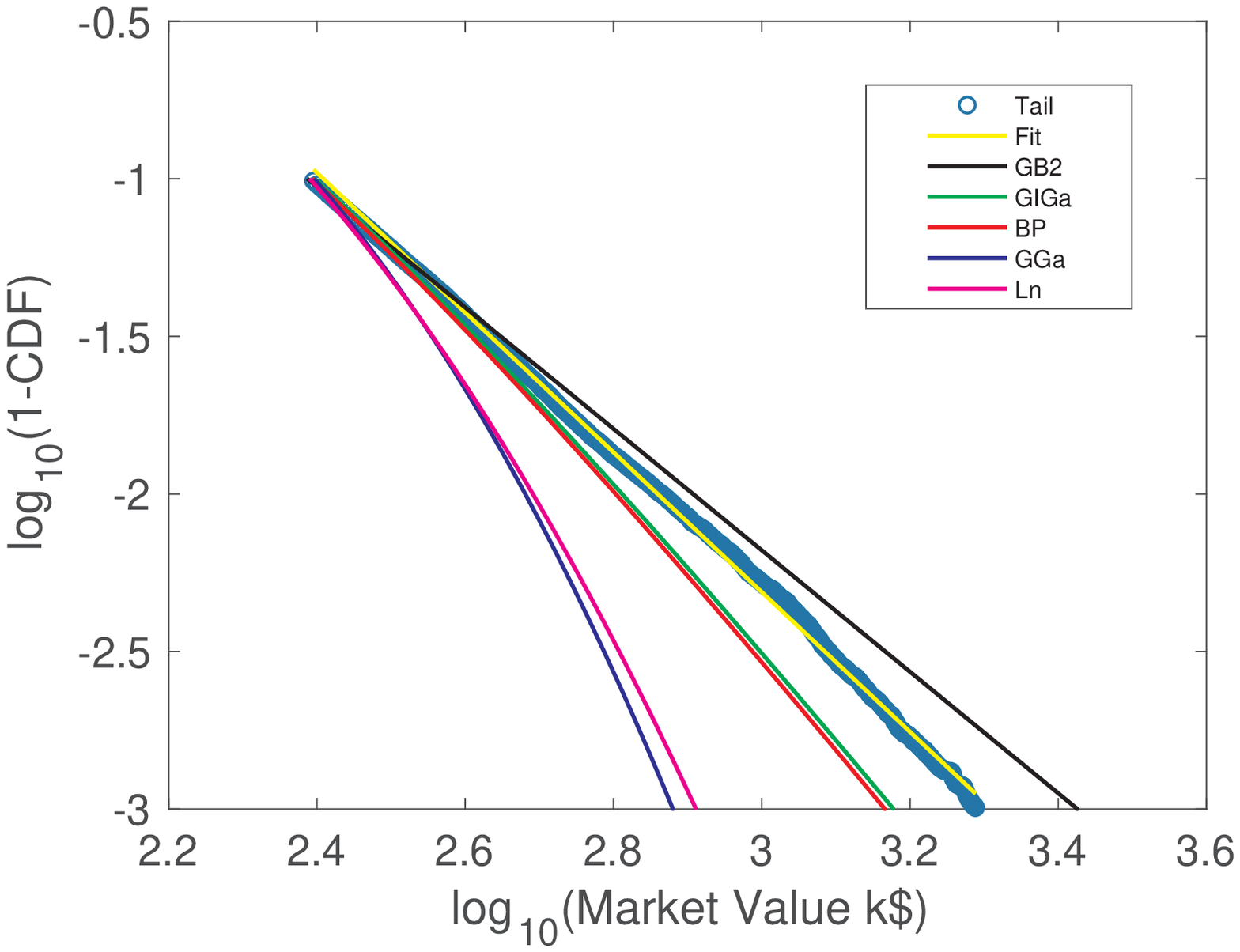}
\end{tabular}
\caption{MLE fits for market values.}
\label{MKT1990}
\end{figure}

It is interesting to compare sale price distribution with that of the original "market value," that is the asking price. We had 124203 data points for the latter. Figs. \ref{MKTSP} and \ref{MKTSPtails} show the contour plots of both distributions, including the top and very top values. Clearly market values have a much "fatter" tail, as confirmed by MLE fits and tail fitting in Fig. \ref{MKT1990} and Tables \ref{MLEMKTTotaldeinf1990} and \ref{TailSlopeMKTTotaldeinf1990}. In fact, the tail exponent is so small as not to allow for theoretical existence of the variance. We believe that this is because there is no underlying model of economic exchange, unlike for sale prices, which are a proxy to wealth/income.

\begin{table}[!htbp]
\centering
\caption{MLE results for market values}
\label{MLEMKTTotaldeinf1990}
\resizebox{\textwidth}{!}{\begin{tabular}{ccccccccccc} 
\hline
            type &       parameters &          KS test &             Mean &              RMS &             Gini &  Hoover     &   Theil T &          Theil L  &  DMMS\\
\hline
Data & N.A.& N.A. &         137.5409 &         169.6051 &           0.4041 &         0.2944 & 0.3419 &           0.2711 &  0.3245\\
GB2 & GB2(1.1477 , 0.5488 , 3.5129 , 70.7373) & 0.0131 & 141.7586 & N.A. & 0.4227 &   0.3054   & N.A. & N.A.&   0.2371\\
BP & BP(         30.9762,           2.9803,           8.5913) &           0.0303 &         134.3902 &           140.0085 &           0.3884 &    0.2804   &    0.2893 &           0.2480 &     0.2156\\
GIGa & GIGa(          3.3516,         340.5329,           0.8940) &           0.0282 &         134.9369 &         143.2419 &           0.3919 &  0.2833      &   0.2959 &           0.2520 &     0.2203\\
LN & LN(          4.6528,           0.6636) &           0.0657 &         130.7109 &          97.2305 &           0.3611 &    0.2599     &  0.2202 &           0.2202 &     0.1722\\
GGa & GGa(         44.4707,           0.00001,           0.2196) &           0.0752 &         131.3837 &          95.9984 &           0.3638 &    0.2620     &  0.2206 &           0.2280 &    0.1594\\
\hline
\end{tabular}}
\end{table}

\begin{table}[!htbp]
\centering
\caption{Tail slope results for market values}
\label{TailSlopeMKTTotaldeinf1990} 
\begin{tabular}{cccc} 
\hline
            type &            Slope & param  & Slope  \\
\hline
Data &    -2.2175 & &\\
GB2  &          -1.9269 & $-q \alpha$ & $-1.928$\\
BP  &          -2.6419 &$-q $ & $-2.980$ \\
GIGa &          -2.6138& $-\alpha \gamma $ & $-2.996$ \\
LN &          -3.9300 & & \\
GGa  &          -4.2277 & & \\
\end{tabular}
\end{table}

\clearpage

\bibliography{mybib}

\begin{thebibliography}{10}
\expandafter\ifx\csname url\endcsname\relax
  \def\url#1{\texttt{#1}}\fi
\expandafter\ifx\csname urlprefix\endcsname\relax\def\urlprefix{URL }\fi
\expandafter\ifx\csname href\endcsname\relax
  \def\href#1#2{#2} \def\path#1{#1}\fi

\bibitem{chotikapanich2008modelling}
D.~Chotikapanjch (Ed.), Modeling Income Distributions and Lorenz Curves,
  Springer, 2008.

\bibitem{mcdonald2008modelling}
J.~B. McDonald, Modeling Income Distributions and Lorenz Curves (Chotikapanich,
  Duangkamon - Editor), Springer, 2008, Ch. 3 and 8.

\bibitem{chen2016influences}
J.~Chen, Y.~Wang, J.~Wen, F.~Fang, M.~Song, The influences of aging population
  and economic growth on chinese rural poverty, Journal of Rural Studies 47
  (2016) 665--676.

\bibitem{chotikapanich2018using}
D.~Chotikapanich, W.~E. Griffiths, G.~Hajargasht, W.~Karunarathne, P.~D.~S.
  Rao, Using the gb2 income distribution, Econometrics 6~(2) (2018) 21.

\bibitem{hertzler2003classical}
G.~Hertzler, "classical" probability distributions for stochastic dynamic
  models, in: 47th Annual Conference of the Australian Agricultural and
  Resource Economics Society, 2003.

\bibitem{bouchaud2000wealth}
J.-P. Bouchaud, M.~M{\'e}zard, Wealth condensation in a simple model of
  economy, Physica A: Statistical Mechanics and its Applications 282~(3) (2000)
  536--545.

\bibitem{ma2013distribution}
T.~Ma, J.~G. Holden, R.~Serota, Distribution of wealth in a network model of
  the economy, Physica A: Statistical Mechanics and its Applications 392~(10)
  (2013) 2434--2441.

\bibitem{dragulescu2002probability}
A.~A. Dragulescu, V.~M. Yakovenko, Probability distribution of returns in the
  heston model with stochastic volatility, Quantitative Finance 2 (2002)
  445--455.

\bibitem{ma2014model}
T.~Ma, R.~Serota, A model for stock returns and volatility, Physica A:
  Statistical Mechanics and its Applications 398 (2014) 89--115.

\bibitem{dashti2018combined}
M.~Dashti~Moghaddam, R.~Serota, Combined mutiplicative-heston model for
  stochastic volatility, arXiv:1807.10793.

\bibitem{mcdonald1995generalization}
J.~B. McDonald, Y.~J. Xu, A generalization of the beta distribution with
  applications, Journal of Econometrics 66 (1995) 133--152.

\bibitem{limpert2001lognormal}
E.~Limpert, W.~Stahel, M.~Abbt, Log-normal distributions across the sciences:
  Keys and clues, BioScience 51~(5) (2001) 341--352.

\bibitem{sarabia2014explicit}
J.~M. Sarabia, V.~Jorda, Explicit expressions of the pietra index for the
  generalized function for the size distribution of income, Physica A 416
  (2014) 582--595.

\bibitem{dashti2019implied}
M.~Dashti~Moghaddam, J.~Liu, R.~A. Serota, Implied and realized volatility: A
  study of distributions and the distribution of difference, arXiv:1906.02306.

\bibitem{yan2019general}
C.~Yan, B.~Zhao, A general jump-diffusion process to price volatility
  derivatives, Journal of Futures Markets 39 (2019) 15--37.

\bibitem{heston1993closed}
S.~L. Heston, A closed-form solution for options with stochastic volatility
  with applications to bond and currency options, The Review of Financial
  Studies 6~(2) (1993) 327--343.

\bibitem{nelson1990arch}
D.~Nelson, Arch models as diffusion approximations\*, Journal of Econometrics
  45 (1990) 7.

\bibitem{fuentes2009universal}
M.~A. Fuentes, A.~Gerig, J.~Vicente, Universal behvior of extreme price
  movements in stock markets, PLoS ONE 4~(12) (2009) 1.

\bibitem{liu2016probability}
Z.~Liu, J.~G. Holden, R.~A. Serota, Probability density of response times and
  neurophysiology of cognition, Advances in Complex Systems 19~(4-5) (2016)
  1650013--1--1650013--17.

\end{thebibliography}

\end{document}